\documentclass{article}
\usepackage[utf8]{inputenc}
\usepackage{authblk}
\usepackage[dvips]{graphicx}
\graphicspath{{noiseimages/}}
\usepackage{xcolor}
\usepackage[T2A]{fontenc}
\usepackage{tikz}
\usepackage{pgfplots}
\usepackage{mathrsfs}
\pgfplotsset{compat=1.7}
\usetikzlibrary{intersections, pgfplots.fillbetween}
\usetikzlibrary{decorations.pathmorphing}
\usetikzlibrary{arrows.meta}
\usepackage[mode=buildnew]{standalone}
\usepackage[toc,page]{appendix}
\usepackage{amsmath}
\usepackage{amssymb}
\usepackage{float}
\usepackage{comment}
\usepackage{a4wide}
\usepackage[english]{babel}
\usepackage{hyperref}
\definecolor{urlcolor}{HTML}{990000}
\definecolor{linkcolor}{HTML}{005F5F}
\hypersetup{pdfstartview=FitH,  linkcolor=linkcolor,urlcolor=urlcolor, colorlinks=true,citecolor=blue}
\usepackage[page]{appendix}

\author[1,2]{E.T.Akhmedov}
\author[1,2]{D.V.Diakonov\footnote{\tt dmitrii.dyakonov@phystech.edu}}

\affil[1]{Institutskii per. 9, Moscow Institute of Physics and Technology, 141700, Dolgoprudny, Russia}
\affil[2]{B. Cheremushkinskaya, 25, Institute for Theoretical and Experimental Physics, 117218, Moscow, Russia}

\title{\textcolor{black}{Free energy and entropy in Rindler and de Sitter space-times}}

\begin{document}

\numberwithin{equation}{section}

\maketitle

\begin{abstract}
We investigate the free energy and entropy of the Gaussian massive scalar field theory in the static de Sitter space-time for arbitrary temperature. For the inverse temperatures of the form $\beta=2 \pi 2^k, \ \ k\in \mathbf{Z}$, in curvature units, we find the explicit form of the free energy and its derivatives with respect to the temperature. There are two types of contributions to the free energy: one is of the ``area type'' and can be attributed to the horizon, while the other is of the ``volume type'' and is associated with the interior of the space-time. The latter contribution in the odd-dimensional case in the limit of the week field (large mass or small Hubble constant) significantly depends on the temperature. Namely, for $ \beta<2\pi$, the free energy behaves as $ F^{bulk}_{\beta} \sim e^{- \beta  \, m} $, while for $\beta>2\pi$ it behaves as $ F^{bulk}_{\beta} \sim e^{- 2 \, \pi \, m}$. We also show that even the leading UV contributions to the free energy significantly depend on the state of the theory, which is very unusual. We explain the origin and physical meaning of these observations. As the model example we consider the situation in the Rindler wedge of the flat space-time.
\end{abstract}
\newpage

\tableofcontents

\newpage

\section{Introduction}

\quad
There are indirect signs that there was a rapid, exponentially expanding epoch in our Universe, which had a de Sitter like geometry \cite{Guth:1980zm, Linde:1981mu, Albrecht:1982wi, Starobinsky:1980te}. Quantum fields in the static patch of the de Sitter space-time are believed to have thermal properties \cite{GH, bm, bgm, sewell, nthir}. 
The canonical Gibbons-Hawking temperature is related to the geometry, i.e., fixed, e.g., by the requirement that the de Sitter metric should be regular when analytically continued to the imaginary time.
This requirement leads to the periodicity of the correlation functions in the Euclidean time coordinate with the period fixed by the inverse Gibbons–Hawking temperature $\beta = 2 \pi$, in the units of the de Sitter curvature.

However, quantum fields in the static de Sitter space-time in the state with the planckian density matrix for exact modes do not possess the same essential properties as those in Minkowksi space-time \cite{Popov:2017xut}. Furthermore, if the temperature is different from the canonical one, the backreaction on the background geometry is strong, as shown in \cite{Akhmedov:2020ryq}, \cite{Akhmedov:2020qxd}, and \cite{Bazarov:2021rrb}. At the end of the day, the thermalization process in the de Sitter space-time, particularly, and in curved space-times, in general, is far from being well understood \cite{Akhmedov:2021rhq}.

We will not consider these problems in the present paper, but we would like to better understand the properties of the thermal states (with the plankian distribution of generic temperature for the exact modes  in the curved space-time). For that reason, in this paper, we calculate the one-loop free energy and entropy of the Gaussian massive scalar field theory in the static de Sitter universe. 

Our calculations extend considerations of earlier works of different authors on the related issues. See e.g. \cite{Dowker:1975tf, Fursaev:1993hm, Das:2006wg,  Maldacena:2012xp, Higuchi:2018tuk} and \cite{Mathur:2021ial}. For generic values of space-time dimension $d$ and inverse temperature $\beta$, we provide an expression for the free energy in terms of an integral. This integral representation agrees with the answer obtained in a recent work \cite{Anninos:2020hfj}, where the free energy in the de Sitter space-time was expressed in terms of the Harish-Chandra character. We discovered this result when we finished work on our paper. 

On top of obtaining the same formula, we find the explicit form of the one-loop effective action (free energy) in the static de Sitter universe of two dimensions for a sequence of temperatures of the form $\beta=2 \pi 2^k; \ \ k\in \mathbf{Z}$. The same sort of explicit form can be found in any dimensions. Then, we study the properties of the free energy in various limits and explain the origin and the physical meaning of the new divergences (as compared to the Minkowski space-time) in the free energy. Then, we calculate the corresponding entropy.

The entanglement entropy contains both UV divergent and finite terms:
\begin{align}
\mathbf{S} = \mathbf{S}_{\text{UV}}+\mathbf{S}_{\text{fin}}.
\end{align}
The general analysis with the use of the heat kernel method in curved space-times with conical singularities \cite{Fursaev:1994in, Casini:2006hu} shows that the structure of the UV divergent terms in $d$ dimensions has the following form:
\begin{align}
\label{2}
\mathbf{S}_{\text{UV}} =  \frac{a_{d-2}}{\epsilon^{d-2}}+...+ \frac{a_{1}}{\epsilon}+a_0\log(\epsilon),
\end{align}
where $\epsilon$ is the short distance cutoff, and $a_i$ are some constants that depend on the geometry of the space-time and physical quantities of the field theory. The coefficient $a_{d-2}$ of the leading divergent term is proportional to the area of the horizon. This sort of behavior of the entropy is usually referred to as the area law \cite{Bek1,Bek2}. We obtain a similar structure in the static de Sitter space-time. As compared to the previous works on the same subject we do all the calculations in the Lorentzian signature and relate the appearance of the extra (area type) terms in the entropy to the peculiarities of the propagators on the horizon. One should keep in mind that the same peculiarities of the propagators lead to the strong backreaction on the geometry \cite{Bazarov:2021rrb}.


The paper is organized as follows. In section 2, we discuss the method of calculation of the effective action or the free energy, using the analytical continuation in complex time. We do that for the free massive scalar field in general static curved space-time, which allows one to express the effective action via the Feynman propagator. In section 2, we use this method for the two simplest cases: in the flat space-time in the Minkowskian and Rindler coordinates. Then we discuss how an extra divergence in the free energy in the Rindler coordinates is associated with the presence of the horizon. In section 3, we discuss the geometry of the static chart of the de Sitter space-time and the canonical quantization of the massive scalar field theory in such a background. Then we construct the thermal Feynman propagator in the theory under consideration. In section 4, we derive an integral representation for the one-loop effective action of the massive scalar field in the two dimensional de Sitter space-time for an arbitrary temperature. We analytically continue the obtained expression to the Euclidian time to obtain the free energy from the effective action. Then we discuss a method of computation of the explicit form of the free energy and its derivatives with respect to the temperature for inverse temperatures from the sequence $\beta=2 \pi 2^k; \ \ k\in \mathbf{Z}.$ In section 5, we extend our considerations to higher dimensional cases and consider various limits and general patterns in the behavior of the free energy in odd space-time dimensions, where the computations are simpler. In section 6, we make conclusions.


\section{The method and setup}

There are several different methods to calculate the free Bose gas partition function in curved space-times, e.g., in one of the standard methods one uses the heat kernel regularization procedure (see e.g. \cite{Vassilevich:2003xt}). In this section, we discuss another approach, in which one uses real-time formalism and then analytically continues to the Euclidian signature.

To show how it works, we apply our method to the massive Gaussian scalar field in $d$ dimensional flat space-time first in the Minkowskian and then in Rindler coordinates. Then, in the next section, we apply this method to the same theory in the de Sitter space-time. We use the flat space-time example just as a toy model or testing ground for our method.

As we shall see, after the analytical continuation to the Euclidian signature, our results for the Rindler space-time agree with those obtained earlier by the standard methods \cite{Kabat:1995eq}.

The effective action for the Gaussian theory is defined as:
\begin{align}
\label{eff}
Z=e^{i \mathcal{S}_{\textsl{eff}}}=\int d [ \varphi] e^{i S[\varphi]} =\int d [\varphi] \, \exp\left[ \frac{i}{2}\int d^d x \sqrt{g}\Big(\partial_\mu \varphi\partial_\mu \varphi-m^2 \varphi ^2 \Big) \right] \propto \det\left[-\Box_g+m^2\right]^{-\frac{1}{2}}.
\end{align}
Rather than calculating \eqref{eff} with the use of the heat kernel method, we instead express its derivative with respect to $m^2$ via the propagator \cite{Dowker:1975tf,Candelas:1975du}. Namely, it is straightforward to see that
\begin{align}
\frac{\partial}{\partial m^2} \log \int d [ \varphi] e^{i S[\varphi]} =- \frac{i}{2} \frac{ \int d^dx \sqrt{-g}\int d [ \varphi] \varphi(x) \varphi(x) e^{i S[\varphi]}}{\int d [ \varphi] e^{i S[\varphi]}  }=-\frac{i}{2} \int d^dx\sqrt{-g}\ G (x,x).
\end{align}
This allows one to express the effective action via the Feynman propagator in the coincidence limit:
\begin{align}
\mathcal{S}_{\textsl{eff}}=-\frac{1}{2}\lim_{M\rightarrow \infty}\int d^dx\sqrt{-g}\int_{M^2}^{m^2} d\bar m^2  \ G (x,x).
\label{Seff}
\end{align}
The coincidence limit should be taken along spatial directions to avoid ambiguities due to the cut in the complex plane of the geodesic distance.

In the case of the finite temperature field theory, one has to use the thermal Feynman propagator:
\begin{align}
G _\beta (x,y)=\frac{\text{Tr} \left[ { e^{-\beta {H}} } \, T \varphi(x)  \varphi(y)  \right] }{\text{Tr}  \ { e^{-\beta {H}}}}.\label{thermalaverage}
\end{align}
where $\beta$ is the inverse temperature. Obviously, the answer for the effective action (or the free energy, after the analytical continuation) depends on the state one chooses for calculating the propagator.

In all, the problem reduces to the construction of the scalar field propagator, then taking the coincidence limit, with an appropriate regularization to be discussed below, and then making the proper analytical continuation to the Euclidian time.

\subsection{Free energy in the Minkowskian coordinates}

Let us start with the simplest case for the thermal effective action for the massive scalar field in flat space-time in the Minkowskian coordinates. This is just a textbook example.

The thermal Feynman propagator in such a case is:
\begin{align}
G_\beta(x,t)=\int \frac{d^{d-1}k}{(2 \pi)^{d-1}} \left[\frac{e^{i \omega_k|t| -i \vec{k}\vec{x}}}{2\omega_k } \frac{1}{e^{\beta \omega_k }-1}+\frac{e^{-i \omega_k |t| +i \vec{k}\vec{x}}}{2\omega_k }\left(1+\frac{1}{e^{\beta \omega_k }-1}\right)\right],
\end{align}
where $\omega_k = \sqrt{\vec{k}^2 + m^2}$.

Then, using \eqref{Seff}, one obtains the effective action of the form:
\begin{align}
\label{Meff}
\mathcal{S}^\beta_{\textsl{eff}}=-\frac{V_{d-1}T}{2}\int_{M^2}^{m^2} d\bar m^2 \int \frac{d^{d-1}k}{(2 \pi)^{d-1} \, \omega_k } \left[\frac{1}{2}+\frac{1}{e^{\beta \omega_k }-1}\right],
\end{align}
where $T$ is the duration of time and $V_{d-1}$ is the spatial volume. The first term under the integral on the right-hand side (RHS) of the last equation leads to the standard UV divergence due to the zero-point fluctuations.
It can be either subtracted or absorbed into the renormalization of the ground state energy after the regularization by the momentum cutoff. The remaining expression remains finite even if one takes $M\to \infty$.

To obtain the free energy, one should perform the Wick rotation $t = - i  \tau$ to the Euclidean signature in (\ref{Meff}). Here one should keep in mind that the thermal Feynman propagator is analytic in the complex $t$-plane on the infinite stripe along the real axis. The stripe is of the width $[0,\beta)$ along the imaginary axis, where $\beta$ is the inverse temperature. This fact is related to the so-called KMS property (see e.g. \cite{LQP}). Due to these analytic properties of the propagator, after the Wick rotation, one should substitute $T \to - i\beta$.  Thus the effective action \eqref{Meff} transforms into:
\begin{align}
\label{FMIN}
i \mathcal{S^\beta}_{\textsl{eff}}\rightarrow  -\beta F_\beta=-\frac{V_{d-1}\beta}{2}\int_{M^2}^{m^2} d\bar m^2 \int \frac{d^{d-1}k}{(2 \pi)^{d-1}}  \frac{1}{\omega_k (e^{\beta \omega_k }-1)}.
\end{align}
This expression is finite and can be transformed into:
\begin{align}
\label{Fmin}
 F_\beta=\frac{V_{d-1}}{\beta}\int \frac{d^{d-1}k}{(2 \pi)^{d-1}} \log\Big[1-e^{-\beta \omega_k}\Big] .
\end{align}
To obtain this expression, we calculated the integral over $\bar{m}$ and took the limit $M\rightarrow \infty$.
For the sake of completeness, let us examine the behavior of this free energy in various limits. In the massless limit, one obtains the well known explicit Stefan-Boltzmann law:
\begin{align}
\label{FMIN1}
F_\beta = \frac{V_{d-1}}{\beta}\frac{S_{d-1}}{(2 \pi)^{d-1}} \int_0^{\infty} d k k^{d-2} \log\left[1-e^{-\beta k}\right ] = - \frac{ V_{d-1} S_{d-1}}{(2 \pi)^{d-1}} \frac{\Gamma (d-1) \zeta (d)}{\beta^{d}}=- \frac{ V_{d-1}}{\beta^d} \frac{\zeta (d) \Gamma\left(\frac{d}{2}\right)}{\pi^{\frac{d}{2}}}.
\end{align}
At the same time, in the limit $\beta m \gg 1$, one finds that:
  \begin{gather}
  \nonumber
F_\beta
 \approx
 -\frac{V_{d-1}}{\beta}\frac{S_{d-1}}{(2 \pi)^{d-1}} \int_0^{\infty} d k k^{d-2} e^{-\beta \sqrt{k^2+m^2}}
 =
 -\frac{V_{d-1}}{\beta}\frac{S_{d-1}}{(2 \pi)^{d-1}} \frac{2^{\frac{d-2}{2}} m^{\frac{d}{2}} \Gamma \left(\frac{d-1}{2}\right)}{\beta
   ^{\frac{d-2}{2}} \sqrt{\pi }} K_{\frac{d}{2}}(\beta m)
   \approx \\ \approx
   -\frac{V_{d-1}}{(2 \pi)^{\frac{d-1}{2}}}\frac{ m^{(d-1)/2}  e^{-\beta m}}{\beta^{(d+1)/2}}.
\label{FMIN2}
\end{gather}
We will use below these limiting expressions to compare with the free energy in Rindler and de Sitter space-times.

\subsection{Free energy in the Rindler chart}

Now let us elaborate our approach in the $d$ dimensional Rindler chart (or wedge, aka quadrant) of flat space-time. The metric of the flat space-time in these coordinates is:
\begin{align}
\label{metricRindler}
    ds^2=e^{2 \xi}\Big(-d\eta^2+d\xi^2\Big) + d\vec{x}^2,
\end{align}
where $\vec{x}$ are the $d - 2$ flat transverse spatial directions. The massive scalar modes in this metric are expressed via the MacDonald functions (see, e.g., \cite{Takagi:1984cd}). The thermal Feynman propagator of the free massive scalar field is \cite{Unruh:1976db, Takagi:1984cd}:
\begin{gather}
\label{RindlerGD}
     G_{\beta}\Big(\eta_2,\xi_2,\vec{x}_2 | \eta_1, \xi_1,\vec{x}_1\Big)
     =\\=
     \nonumber
     \int \frac{d^{d-2}k}{(2\pi)^{d-2}} \int_{0}^{\infty}  \frac{d\omega}{\pi^2}\Bigg[  e^{i \omega |\eta_2-\eta_1|-i\vec{k}\Delta \vec{x}}  \sinh(\pi \omega)  K_{i\omega}\Big(\sqrt{m^2+k^2} e^{\xi_1}\Big) \, K_{i\omega}\Big(\sqrt{m^2+k^2} e^{\xi_2}\Big) \frac{1}{e^{\beta \omega}-1}
          +\\+
               \nonumber
         e^{-i \omega|\eta_2-\eta_1|+i\vec{k}\Delta \vec{x}}   \sinh(\pi \omega)  K_{i\omega}\Big(\sqrt{m^2+k^2} e^{\xi_1}\Big) \, K_{i\omega}\Big(\sqrt{m^2+k^2} e^{\xi_2}\Big) \left(1+\frac{1}{e^{\beta \omega}-1}\right) \Bigg].
\end{gather}
After the substitution of \eqref{RindlerGD} into \eqref{Seff} and integration over the $d-2$ transverse spatial directions and time, one obtains that:
\begin{gather}
\label{EffRinKK}
\mathcal{S^\beta}_{\textsl{eff}}=  - T A_{d-2}\int_{M^2}^{m^2} d\bar m^2 \int \frac{d^{d-2}k}{(2\pi)^{d-2}}  \int_{0}^{+\infty} \frac{d\omega}{\pi^2} \sinh(\pi \omega) \left[\frac12+\frac{1}{e^{\beta \omega}-1}\right]
\times  \\ \times \nonumber
\int_{-\infty}^\infty d \xi e^{2\xi} K_{i\omega}\Big(\sqrt{\bar m^2+k^2} e^{\xi}\Big) \, K_{i\omega}\Big(\sqrt{\bar m^2+k^2} e^{\xi}\Big),
\end{gather}
where $A_{d-2}$ is the volume of the transverse $(d - 2)$-dimensional flat space, and $T$ is the duration of time. This expression has several divergences.
The first divergence is coming from the $1/2$ term under the square brackets in the integral over $\omega$. This is the standard UV divergence due to the zero-point fluctuations. The divergence is independent of the temperature and is similar to the one in the Minkowski space-time \eqref{Meff}. So we treat it the same way as in the previous subsection, i.e., subtract it via renormalization. The remaining divergence we cutoff by restricting the integration over the momenta by $\left|\vec{k}\right|\leq \Lambda$. Such a divergence is not present in the Minkowski coordinates. We elaborate on its nature in the next subsection.

Then, using the relation \cite{Grad}:

\begin{align}
\label{kktabint}
K_{i\omega}\Big(\sqrt{\bar m^2+k^2} e^{\xi}\Big) \, K_{i\omega}\Big(\sqrt{\bar m^2+k^2}e^{\xi}\Big) =2 \int_0^\infty d \lambda \  K_0\left(2\sqrt{\bar m^2+k^2} e^{\xi}\cosh(\lambda)\right) \cos(2\lambda \omega ),
\end{align}
one can take the integral in \eqref{EffRinKK} over $\xi$ and $\bar m^2$ to find that:
\begin{gather}
\mathcal{S^\beta}_{\textsl{eff}}=  -\frac{T A_{d-2}}{2}\int\limits^{\left|\vec{k}\right|\leq \Lambda}\frac{d^{d-2}k}{(2\pi)^{d-2}}  \log\frac{ k^2+ m^2}{k^2+M^2} \int_{0}^{+\infty}  \frac{d\omega}{\pi^2} \frac{\sinh(\pi \omega)}{e^{\beta \omega}-1} \, \int_0^\infty d \lambda \ \frac{ \cos(2\lambda \omega ) }{ \cosh^2(\lambda)}
\label{EffRindBeta}
\end{gather}
Then, using the standard integral representation of the logarithm:
$$
\log{\frac{B}{A}}=\int_0^\infty \frac{ds}{s}\left(e^{-s A}-e^{-s B}\right),
$$
one obtains that:
\begin{gather}
\mathcal{S^\beta}_{\textsl{eff}}=  \frac{T A_{d-2}}{2}\int\limits^{\left|\vec{k}\right|\leq \Lambda}\frac{d^{d-2}k}{(2\pi)^{d-2}}  \int_0^\infty \frac{ds}{s}\left(e^{-s ( k^2+ m^2)}-e^{-s (k^2+M^2) }\right) \times \\ \nonumber
\times \int_{0}^{+\infty}  \frac{d\omega}{\pi^2} \frac{\sinh(\pi \omega)}{e^{\beta \omega}-1} \, \int_0^\infty d \lambda \ \frac{ \cos(2\lambda \omega ) }{ \cosh^2(\lambda)}.
\end{gather}
In this expression we can cutoff the lower limit of integration over $s$ by $\epsilon^2 \to 0$. This allow one to take $\Lambda \to \infty$. Then all the expressions remain finite. After that one can take the Gaussian integrals over $k$ to obtain that:
\begin{align}
\label{Srindc}
\mathcal{S^\beta}_{\textsl{eff}}=   \frac{2 T A_{d-2}}{\pi}  \int_{0}^{+\infty} d\omega \frac{\sinh(\pi \omega)}{e^{\beta \omega}-1} \int_0^\infty d \lambda \frac{ \cos(2\lambda \omega )}{\cosh^{2}(\lambda)}
\int_{\epsilon^2}^\infty \frac{  d s  }{\left( 4 \pi s\right)^{\frac{d}{2}} }    \left(e^{  -m^2 s }-e^{  -M^2 s }\right).
\end{align}
Taking the integral over $\lambda$ in the last expression we obtain the effective action of the form\footnote{For the massless theory in $d=2$ dimensions, the $s$ integral in (\ref{entRind}) is also divergent in the IR limit, i.e., the mass $m$ must be kept to be non-zero to provide the IR cutoff.}:
\begin{gather}
\label{entRind}
\mathcal{S^\beta}_{\textsl{eff}}=
2 T A_{d-2}  \int_{0}^\infty \ d\omega   \frac{ \omega}{e^{\beta \omega}-1} \int_{\epsilon^2}^\infty\frac{ds}{(4\pi s)^{\frac{d}{2}}} \left( e^{-s m^2}-e^{-s  M^2}\right)
\end{gather}
Here one can safely take the limit $M\rightarrow \infty$, because all integrals are finite for non-zero $\epsilon$.

At the same time, in all the equations after \eqref{metricRindler}, one can perform the Wick rotation $t = - i  \tau$ to the Euclidean signature.  Again one should keep in mind\footnote{One should also keep in mind that for generic values of $\beta$, the analytically continued metric \eqref{metricRindler} has the conical singularity with the deficit angle $2 \pi - \beta$. This is relevant for the discussion of the next subsection \cite{Fursaev:1994ea}.} the analytic properties of the thermal Feynman propagator in the complex $t$-plane and substitute $T \to - i \beta$.

In all, after the Wick rotation the effective action \eqref{entRind} transforms into:
\begin{gather}
\label{EFFm}
i \mathcal{S^\beta}_{\textsl{eff}}\rightarrow  -\beta F_\beta =
\beta A_{d-2}   \int_{0}^\infty \ d\omega   \frac{2 \omega}{e^{\beta \omega}-1} \int_{\epsilon^2}^ \infty\frac{ds}{(4\pi s)^{\frac{d}{2}}} e^{-s m^2} .
\end{gather}
Taking the integral over $\omega$ one obtains that the free energy is equal to:
\begin{align}
\label{EFFM}
  F_\beta=
-\frac{ \pi^2  A_{d-2}}{3 }    \frac{1}{\beta^2} \int_{\epsilon^2}^ \infty\frac{ds}{(4\pi s)^{\frac{d}{2}}} e^{-s m^2}.
\end{align}
This is the result that was found earlier in \cite{Kabat:1995eq}. The UV divergent contribution to the entropy that follows from \eqref{EFFM} agrees with \eqref{2}.

In the two dimensional case it is more convenient to restore $M$ and take the limit $\epsilon \rightarrow 0$ to represent the answer for the free energy in the following form:
\begin{align}
F_\beta=\frac{\pi}{6 \beta^2} \log \frac{m}{M}.\label{mMRind}
\end{align}
As one can see now, there is an essential difference between the free energy of the scalar field in the Rindler \eqref{EFFM} and Minkowski \eqref{Fmin} coordinates in the flat space-time. First, while in the Rindler coordinates, the free energy is proportional to the temperature in the second power (and this does not depend on the dimension of the space-time), in the Minkowski space-time, this dependence is essentially related to the space-time dimension (see \eqref{FMIN1} and \eqref{FMIN2}). Second, the free energy in the Minkowski coordinates is proportional to the spacial volume $V_{d-1}$, while in the Rindler coordinates the free energy is proportional to the ''area'' of the horizon $A_{d-2}$. 
Third, after subtracting the zero-point fluctuations, the
free energy in Minkowski coordinates is finite in contrast to the Rindler one.

Apparently, all these differences in the free energy for various charts can be traced back to the fact that in different coordinate systems, we use different modes and Cauchy surfaces of different topology (see, e.g., \cite{Akhmedov:2021rhq} for a related discussion). The presence of the horizon is obviously crucial. Below we discuss these issues in greater details.

\subsection{Discussion of the extra divergence present in the Rindler chart}

In \cite{Akhmedov:2020qxd, Akhmedov:2020ryq}, it was shown that such a thermal propagator, which we used above in the Rindler chart, has an anomalous singularity at the horizon. That happens for non--canonical temperatures $\beta \neq 2\pi$ in the acceleration units.

The simplest idea how to see the presence of the anomalous divergence in question is as follows. Let us examine the behavior of the propagator for $\beta=\pi$ near the horizon. For this temperature, one can use the relation:
\begin{align}
\frac{1}{e^{\pi \omega}-1}=\frac{1}{e^{2\pi \omega}-1}+\frac{e^{\pi \omega}}{e^{2\pi \omega}-1},
\end{align}
and eq. \eqref{RindlerGD}. Then, for $\beta= \pi$ one can represent the propagator in the following form:
\begin{gather}
\label{anomdiv}
G_\pi \Big(\eta_2,\xi_2,\vec{x}_2 | \eta_1, \xi_1,\vec{x}_1\Big)
=\\=
\nonumber
G_{2 \pi} \Big(e^{2\xi_1}+e^{2\xi_2}-2 e^{\xi_1+\xi_2} \cosh(\eta_1-\eta_2)+(\vec{x}_1-\vec{x}_2)^2\Big) 
+\\+
\nonumber
G_{2 \pi} \Big(e^{2\xi_1}+e^{2\xi_2}+2 e^{\xi_1+\xi_2} \cosh(\eta_1-\eta_2)+(\vec{x}_1-\vec{x}_2)^2\Big).
\end{gather}
Here the first term is the standard Poincare invariant two-point function for the canonical temperature $\beta = 2\pi$. It is the same as in Minkowski space-time. It has the standard UV divergence for the light-like separations of its points.

At the same time, the second term in \eqref{anomdiv} is finite inside the Rindler wedge but becomes singular once both its points are taken to the same side of the horizon --- boundary of the wedge. The second term's argument vanishes when both points are light–like separated on the same side of the horizon  (future or past,  $\xi_{1,2} \to \pm \infty$). As a result, when both points of the propagator are light-like separated on the horizon, they has the standard UV divergence but with the wrong (anomalous) coefficient. (For the temperature under consideration corresponding to $\beta = \pi$, the coefficient is wrong by the factor of two.) It can be shown that the coefficient depends on the temperature \cite{Akhmedov:2020qxd, Akhmedov:2020ryq}. This subsection shows how this anomalous divergence manifests itself in the effective action and is related to the discussion above.

Let us again substitute \eqref{RindlerGD} into \eqref{Seff}, but calculate first the integrals over the momenta $k$ instead of those over the transverse $d-2$ spatial directions. To do this calculation, let us represent the equations in a more suitable form. Namely, we use in \eqref{kktabint} of the following representation:
\begin{align}
K_0\left(2\sqrt{m^2+k^2} e^{\xi}\cosh(\lambda)\right)=\int_0^\infty\frac{ d \rho}{2\rho}  \exp\left[    -\rho -\frac{(m^2+k^2) e^{2\xi}\cosh^2(\lambda)}{\rho}\right].
\end{align}
Then instead of \eqref{EffRinKK}, we obtain:
\begin{gather}\label{exp225}
\mathcal{S^\beta}_{\textsl{eff}}=  -T A_{d-2}\int_{M^2}^{m^2} d\bar m^2 \int \frac{d^{d-2}k}{(2\pi)^{d-2}}  \int_{0}^{+\infty} \frac{d\omega}{\pi^2}  \sinh(\pi \omega) \left[\frac{1}{2}+\frac{1}{e^{\beta \omega}-1}\right]
\times  \\ \times \nonumber
\int_{-\infty}^\infty d \xi e^{2\xi}   \int_0^\infty d \lambda \cos(2\lambda \omega )\int_0^\infty\frac{ d \rho}{\rho}  \exp\left[    -\rho -\frac{(m^2+k^2) e^{2\xi}\cosh^2(\lambda)}{\rho}\right].
\end{gather}
Here we treat as usual the divergent contribution from the zero-point fluctuations. But (\ref{exp225}) has also other divergences. In fact, let us change the variables $e^{2\xi} =  s$ in (\ref{hordiv}) and impose the cutoff parameter $\epsilon^2$ at the lower bound of the integration over $s$. This cutoff can be interpreted as a smallest possible distance to the horizon.
Now one can take $M\to \infty$ and one does not need to regularize the integral over $k$. Then, taking the integrals over the momentum and mass in the last expression, one obtains:
\begin{gather}
\label{hordiv}
\mathcal{S^\beta}_{\textsl{eff}}=  -T A_{d-2}  \int_{0}^{+\infty} \frac{d\omega}{\pi^2} \frac{\sinh(\pi \omega)}{e^{\beta \omega}-1}\int_0^\infty d \lambda \frac{ \cos(2\lambda \omega )}{\cosh^{d}(\lambda)}
\times  \\ \times \nonumber
\int_0^\infty d \rho\rho^{\frac{d-2}{2}} e^{-\rho} 2\pi  \int_{\epsilon^2}^\infty \frac{d s  }{\left( 4 \pi s\right)^{\frac{d}{2}} }    \exp\left[  -\frac{m^2 s\cosh^2(\lambda)}{\rho}\right].
\end{gather}
The last integral here is divergent at the horizon of the Rindler coordinates (\ref{metricRindler}), i.e. in the limit $ \epsilon\rightarrow 0$. As we can see now this divergence appears due to the singularity of the thermal propagator at the horizon, which we have described above in this section and which was encountered in \cite{Akhmedov:2020qxd, Akhmedov:2020ryq}.

Let us now rescale the integration variable $s$ in \eqref{hordiv} to obtain:
\begin{gather}
\label{hordiv2}
\mathcal{S^\beta}_{\textsl{eff}}=  -     \frac{2 T A_{d-2}}{\pi}  \int_{0}^{+\infty} d\omega \frac{\sinh(\pi \omega)}{e^{\beta \omega}-1} \int_0^\infty d \lambda \frac{ \cos(2\lambda \omega )}{\cosh^{2}(\lambda)}
\times  \\ \times \nonumber
\int_0^\infty d \rho e^{-\rho} \int_{\epsilon^2\frac{ \rho}{\cosh^2(\lambda)}}^\infty \frac{  d s  }{\left( 4 \pi s\right)^{\frac{d}{2}} }    e^{  -m^2 s }.
\end{gather}
Note that the finite parts, $\epsilon^0$, of the regularized effective actions \eqref{EFFm} and \eqref{hordiv2} do coincide.
At the same time, the divergent parts of \eqref{hordiv2} and \eqref{EFFm} are different. However, if one makes the replacement $\epsilon^2\rightarrow  \epsilon^2\frac{\cosh^2(\lambda)}{\rho}$ in \eqref{hordiv2}, then after the calculation of the $\lambda$ and $\rho$ integrals, one reproduces \eqref{EFFm}.


In all, there is a divergent contribution in the effective action (free energy) in the Rindler coordinates, directly related to the presence of the horizon. There is no such divergence in the Minkowskian coordinates. The divergent contribution is proportional to the horizon area rather than to the volume of space-time. At the same time, to obtain \eqref{hordiv2} we subtracted (renormalized) the divergence due to zero-point fluctuations, which are present even in the ground state, $\beta = \infty$, for the exact modes.
Note that such a divergence as in \eqref{hordiv2} is also present for the canonical temperature, i.e., when the Feynman thermal propagator in the Rindler chart is equivalent to the Poincaré invariant Feynman propagator in the Minkowskian coordinates. 

\section{Quantization in the static chart of the de Sitter space-time}

In this section, we discuss quantization of the massive scalar field in the $d\geq 2$ dimensional static patch of the de Sitter space-time. The case $d=2$ for any value of $\beta$ was considered in detail in \cite{Akhmedov:2020qxd}, while the case $d=4$ only for the canonical temperature $\beta = 2\pi$ was elaborated in \cite{Polarski:1989iu, Polarski:1990tr}.

\subsection{Geometry of the static chart of the de Sitter space-time}

The $d$-dimensional de Sitter space-time is the hyperboloid embedded in the $(d+1)$--dimensional ambient Minkowski space-time:
\begin{align}
dS_d=\{ X  \in \mathbf{R}^{d,1}, \ X_\alpha X^\alpha =-R^2 \}, \quad \alpha = \overline{0,d}.
\end{align}
In what follows, we set the de Sitter radius to $R = 1$. The static patch of the de Sitter space-time is covered by the coordinates as follows:
\begin{equation}
\label{coordinates}
 \begin{cases}
   X^0=\sqrt{1-r^2}\sinh t \\
   X^1= \sqrt{1-r^2} \cosh t  \\
   X^i=r z_i \quad 2 \leq i \leq d
 \end{cases},
 \qquad t \in(-\infty,\infty), \ r \in(0,1).
\end{equation}
Where $z_i$ are the coordinates on the $(d-2)$--dimensional sphere. In these coordinates, the de Sitter metric, induced from the flat one of the ambient space-time, takes the form:
\begin{align}
\label{metricdS}
ds^2=-\left(1-r^2\right) dt^2+\left(1-r^2\right) ^{-1} dr^2+r^2 d\Omega_{d-2}^2.
\end{align}
One of the main properties of this metric is the existence of the time-like Killing vector, which allows one to introduce the notion of energy and define the thermal state with planckian distribution for the exact modes and the inverse temperature $\beta$.

The static patch is bordered by the Killing horizon at $r^2=1$, where the metric degenerates. Hence, the aforementioned time-like killing vector is not globally defined. Note that in a two-dimensional de Sitter space-time, the range of the $r$ coordinate is different: $ r \in(-1,1)$; $2d$ static patch is bordered by the bifurcate Killing horizon.

\subsection{Quantization}

Now let us apply the standard method of the canonical quantization to the Gaussian massive scalar field theory on the background \eqref{metricdS}. The Klein–Gordon equation for the field is as follows:
\begin{align}
(\Box+m^2)\phi=\left[\frac{\partial_t^2}{1-r^2}-\frac{1}{r^{d-2}}\partial_r\Big(r^{d-2}(1-r^2)\partial_r\Big) - \frac{\triangle_{\Omega}}{r^2} + m^2\right]\,\phi=0. \label{KGeq}
\end{align}
Here $\triangle_{\Omega}$ is the Laplace operator on the unit sphere. The positive frequency mode is defined by the separation variables as:
\begin{align}
\phi_{\omega j k}(t,r,\Omega)=\frac{1}{\sqrt{2 \omega}}e^{-i \omega  t} R_{\omega j}(r) Y_{j k}(\Omega). \label{phiRY}
\end{align}
The hyperspherical harmonics $Y_{j k}(\Omega)$ solve the following equation:
\begin{align}
\triangle_{\Omega}  Y_{j k}(\Omega)=-j(j+d-3) Y_{j k}(\Omega),
\end{align}
where $j$ is the non-negative integer, $k = (1, 2, .., N_{j,d})$ is the multi-index and the degeneracy of these harmonics $N_{j,d}$ is given by:
\begin{align}
 N_{j,d}=\frac{(j+d-4)!}{(d-3)! j!}(2j+d-3),
\end{align}
with the exceptional case $d=3$ and $j=0$, where $N_{0,3}=1$. The radial function $R_{\omega j}$ solves the equation:
\begin{align}
\label{Rdifeq}
\left(-\frac{\omega^2}{1-r^2}-\frac{1}{r^{d-2}}\partial_r\left(r^{d-2}(1-r^2)\partial_r\right)+\frac{j(j+d-3) }{r^2}+m^2\right)R_{\omega j}(r)=0,
\end{align}
as follows from (\ref{KGeq}) and \eqref{phiRY}.

The last equation has two linearly independent solutions. In the two-dimensional case, these solutions are Ferrers functions $\mathsf{P}^{i \omega}_{-\frac12 +i \nu}(\pm r) $, also known as Legendre functions on the cut. Both are regular and normalized as follows \cite{Akhmedov:2020qxd}:
\begin{align}
R_{\omega}(\pm r)=\frac{\Gamma\left(\frac{1}{2}+i \nu-i \omega\right)\Gamma\left(\frac{1}{2}-i \nu-i \omega\right)}{\sqrt{2 \pi}\Gamma(i\omega)} \mathsf{P}^{i \omega}_{-\frac12 +i \nu}(\pm r),
\end{align}
where
\begin{equation}
    \nu =\sqrt{m^2-\left(\frac{1}{2}\right)^2}.
\end{equation}
The mode expansion of the field operator can then be written as follows:
\begin{align}
\label{harmexp2}
\phi (t,r)=\int_0^\infty d\omega\Big[ \phi_{\omega }(t,r) a_{\omega }+\phi_{\omega }(t,-r) b_{\omega }+\phi^*_{\omega }(t,r) a^\dagger_{\omega }+\phi ^* _{\omega }(t,-r) b^\dagger_{\omega }  \Big]
\end{align}
Here, the double degeneracy of the energy level $\omega$ appears due to the presence of the left and right moving modes. As a result, one has to introduce two pairs of creation and annihilation operators
for each level:
\begin{align}
[a_{\omega },a^\dagger_{\omega' }]=\delta(\omega-\omega') , \quad [b_{\omega },b^\dagger_{\omega' }]=\delta(\omega-\omega') \quad \text{and} \quad [a_{\omega },b_{\omega' }]=0.
\end{align}
At the same time, in the case when the dimension of space-time is more than two, only one linear combination of the two independent solutions is regular\footnote{Note that Equation (\ref{Rdifeq}) does have a singularity at $r = \pm 1$ and at $r=0$, but normalization condition does not impose any regularity conditions at $r=\pm 1$.
Meanwhile, the normalization condition does impose the regularity condition at $r=0$ in $d>2$.} at $r=0$:
\begin{align}
\label{Rmode}
R_{\omega j}(r)=A_{\omega j} r^j (1-r^2)^{\frac{i\omega }{2}}      {}_2 F_1\left(\frac{ i \omega - i \nu+ j+\frac{d-1}{2}}{2},\frac{ i \omega + i \nu+ j+\frac{d-1}{2}}{2}, j+\frac{d-1}{2},r^2\right),
\end{align}
where $_2F_1 $ is the hypergeometric function, $A_{\omega j}$ is some normalization constant to be defined below and
\begin{equation}
    \nu =\sqrt{m^2-\left(\frac{d-1}{2}\right)^2}.
    \label{nunu}
\end{equation}
Note that in two dimensions, $r \in (-1,1)$, while in $d>2$, the range is $r \in [0,1)$ and $r=0$ is not the boundary.

Then for the case when $d>2$, the mode expansion of the field operator is as follows\footnote{The way of quantization we are applying here is adequate in such a symmetric space as de Sitter. We use here the modes which have a fixed form in the entire space-time. In fact, the tree-level propagator constructed for the Fock space ground state with the use of such a mode expansion as (\ref{harmexp}) is a function of the geodesic distance between its points. There is, however, a different way to quantize in any space-time: one has to choose a Cauchy surface and a basis of mode functions (of only spatial coordinate) and then evolve the corresponding field operator with the use of the free Hamiltonian. However, the latter way of quantization does not respect the de Sitter isometry: already tree-level propagator does depend separately on each of its arguments. That is the reason why we choose the first method to quantize the theory in de Sitter space-time.}:
\begin{align}
\phi(t,r,\Omega)=\int _0^\infty d\omega\sum_{j,k} \left(\phi_{\omega j k} a_{\omega j k}+\phi^*_{\omega j k} a^\dagger_{\omega j k}\right),
\label{harmexp}
\end{align}
where the creation and annihilation operators obey the canonical commutation relations:
\begin{align}
\label{comrelat}
[a_{\omega j k},a^\dagger_{\omega' j' k'}]=\delta(\omega-\omega')\delta_{j,j'}\delta_{k,k'} \quad\text{and}\quad [a_{\omega j k},a_{\omega' j' k'}]=0.
\end{align}
The vacuum is defined as the state that is annihilated by all annihilation operators: $a_{\omega j k}|0\rangle=0$.

To fix the normalization constant $A_{\omega j }$, we demand that the field operator and its conjugate momentum should obey the canonical commutation relations:
\begin{align}
\label{concomrel}
\Big[\phi(t,r,\Omega),\partial_t\phi(t,r',\Omega')\Big]=\frac{i}{\sqrt{g}g^{tt}}\delta(r-r')\delta(\Omega-\Omega').
\end{align}
Using the completeness relation for the hyperspherical harmonics $Y_{j k}(\Omega)$:
\begin{align}
\sum_{j=0}^{\infty}\sum_{k=0}^{N_{j,d}} Y_{j k}(\Omega) Y_{j k}(\Omega')=\frac{1}{\sqrt{g_\Omega}} \delta(\Omega-\Omega')
\end{align}
and \eqref{comrelat}, one can reduce the relation \eqref{concomrel} to the completeness condition for the basis of radial functions:
\begin{align}
\label{comrelR}
\int_0 ^\infty d \omega  R_{\omega j}(r) R^*_{\omega j}(r')  =
\frac{1-r^2}{r^{d-2}}\delta(r-r').
\end{align}
Instead of finding the normalization from the completeness relation, it is much easier to find it from the orthogonality condition for the radial components of the modes. In fact, multiplying both sides of \eqref{comrelR} by $\frac{r'^{d-2}}{1-r'^2}  R_{\omega j}(r')$ and integrating over $r'$, it is easy to see that the orthogonality condition should be of the following form:
\begin{align}
\label{Rort}
\int_0^1 dr \frac{r^{d-2}}{1-r^2}   R_{\omega j}(r) R^*_{\omega' j}(r)  =
\delta(\omega-\omega').
\end{align}
Using eq. \eqref{Rdifeq}, one can show that the integrand on the left hand side (LHS) of \eqref{Rort} is the total derivative:
\begin{gather}
\nonumber
I_{\omega \omega'}=\int_0^1 dr \frac{r^{d-2}}{1-r^2}   R_{\omega j}(r) R^*_{\omega' j}(r)
=\\=
\nonumber
\frac{1}{\omega'^2-\omega^2}
\int_0^1 dr \Big[ R_{\omega j}(r) \partial_r\Big(r^{d-2}(1-r^2)\partial_r  R^*_{\omega' j}(r) \Big)-R^*_{\omega' j}(r) \partial_r\Big(r^{d-2}(1-r^2)\partial_r  R_{\omega j}(r) \Big)\Big]
=\\=\frac{r^{d-2}(1-r^2)}{\omega'^2-\omega^2}
 \Big[ R_{\omega j}(r) \partial_r  R^*_{\omega' j}(r) - R^*_{\omega' j}(r)\partial_r  R_{\omega j}(r)\Big]\bigg|^{1}_0.
 \label{profort}
\end{gather}
The contribution at the lower limit of this integral is vanishing since the radial function is regular at the origin. To find the contribution from the upper limit, one can use the asymptotic behavior of $R_{\omega j}(r)$ near $r=1$:


\begin{align}
\label{asybeh}
R_{\omega j }(r) \approx
A(\omega)  (2-2 r)^{-\frac{i w}{2}}+B(\omega)  (2-2 r)^{\frac{i w}{2}},
\end{align}
where
\begin{align}
A(\omega)=  A_{\omega j } \frac{ \Gamma (i w)\Gamma \left(\frac{d-1}{2}+l\right)}{ \Gamma \left(\frac{d+2 l-2 i \nu +2 i w-1}{4} \right) \Gamma \left(\frac{d+2 l+2 i \nu +2 i w-1}{4} \right)}
\end{align}
and
\begin{align}
B(\omega)= A_{\omega j } \frac{\Gamma (-i w)\Gamma \left(\frac{d-1}{2}+l\right)}{\Gamma \left(\frac{d+2 l-2 i \nu -2 i w-1}{4}\right) \Gamma \left(\frac{d+2 l+2 i \nu
   -2 i w-1}{4}\right)}.
\end{align}
Using the asymptotic behavior \eqref{asybeh}, one obtains that:
\begin{gather}
\nonumber
I_{\omega \omega'}=\lim_{r\to 1}\left[-\frac{i  (2-2 r)^{-\frac{1}{2} i \left(w'+w\right)}
}{w'+w}  A(w)  B\left(w'\right)^*-\frac{i  (2-2 r)^{-\frac{1}{2} i \left(w-w'\right)}
  }{w-w'} A(w) A\left(w'\right)^*
  +\right. \\ \left. +\frac{i (2-2 r)^{\frac{1}{2} i \left(w'+w\right)}
}{w'+w}  B(w)   A\left(w'\right)^*+\frac{i  (2-2 r)^{\frac{1}{2} i \left(w-w'\right)}
}{w-w'} B(w)   B\left(w'\right)^* \right].
\label{Ilim}
\end{gather}
Then, in \eqref{Ilim}, we use:
\begin{gather}
\label{limdelta}
\lim_{r\rightarrow1} \frac{(2-2 r)^{\frac{1}{2} i \left(w\pm w'\right)}}{w\pm w'}
=\\=
\lim_{r\rightarrow1}\Big(\frac{\cos \left(\frac{1}{2} \log (2-2 r)
   \left(w\pm w'\right)\right)}{w\pm w'}+\frac{i \sin \left(\frac{1}{2}
   \log (2-2 r) \left(w\pm w'\right)\right)}{w\pm w'}\Big)
   = -i \pi \delta(w\pm w').
   \nonumber
\end{gather}
Here we also use one of the standard representations of the delta-function and the Riemann–Lebesgue lemma, which states that
$$
\lim_{r\rightarrow 1} \frac{\cos \left(\frac{1}{2} \log (2-2 r)
   \left(w\pm w'\right)\right)}{w\pm w'}=0
$$
in the distributional sense.

After that, taking into account eq. \eqref{limdelta}, we obtain:
\begin{align}
I_{\omega \omega'}=\pi \left(|A(\omega )|^2+|B(\omega) |^2 \right) \delta(\omega-\omega').
\end{align}
As a result, to obtain proper commutation relations, the normalization constant of the modes (\ref{Rmode}) should be defined as:
\begin{align}\label{Aomj}
|A_{\omega j}|^2=\frac{ |\Gamma \left(\frac{d+2 l-2 i \nu -2 i w-1}{4} \right) \Gamma \left(\frac{d+2 l+2 i \nu -2 i w-1}{4}\right) |^2}{2 \pi |\Gamma \left( i \omega\right)| ^2\, \Gamma^2\left(\frac{d-1}{2}+l\right)}.
\end{align}
This concludes the discussion of the canonical quantization in the static chart of the de Sitter space-time.

\subsection{The thermal propagator}
Now we find the thermal Wightman function since it is the building block that is needed to find the Feynman propagator and, then, to calculate the effective action.

The expectation value over the quantum thermal state with the inverse temperature $\beta$ is defined as in \eqref{thermalaverage},
where the free Hamiltonian in the static chart of the de Sitter space-time is given by:
\begin{gather}
 \label{hamiltionian}
 {H} := \int dx \, \sqrt{g} \,  T^0_{\,0} =\int_0^{+\infty} d\omega \ \omega \  \left(  {a}_{\omega}^{\dag}{a}_{\omega }+ {b}_{\omega }^{\dag}{b}_{\omega}\right)\quad \text{for} \quad  d=2  \quad\text{and}
 \\
\nonumber
{H} := \int d^{d-1}x \, \sqrt{g} \,  T^0_{\,0} = \sum_{j,k} \int_0^{+\infty} d\omega \ \omega \   {a}_{\omega j k}^{\dag}{a}_{\omega j k}\quad \text{for} \quad  d>2.
 \end{gather}
Here, $T^0_{\,0}$ is the corresponding component of the stress-energy tensor of the theory (\ref{eff}) in the metric (\ref{metricdS}). To obtain the expressions for the Hamiltonians, we used the mode expansions (\ref{harmexp2}) and (\ref{harmexp}) with the modes found in the previous subsection.

Then the corresponding thermal Wightman function is:
\begin{gather}
\nonumber
W_\beta\Big(t_1-t_2,\, r_1,r_2\Big) = \Big< {\phi}(t_1,r_1){\phi}(t_2,r_2) \Big> =\\= \int_{0}^{\infty}d\omega  d\omega'   \Bigg[\phi_{\omega }(t_1,r_1)\phi^*_{\omega' }(t_2,r_2) \Big<  a_{\omega }  a_{\omega' }^{\dag} \Big> +\phi_{\omega }(t_1,-r_1)\phi^*_{\omega' }(t_2,-r_2) \Big<  b_{\omega }  b_{\omega' }^{\dag} \Big>
\nonumber
\label{wightman02}
+\\+
\phi^*_{\omega }(t_1,r_1)\phi_{\omega'}(t_2,r_2) \Big<  a_{\omega }^{\dag}   a_{\omega' }\Big>+\phi^*_{\omega }(t_1,-r_1)\phi_{\omega'}(t_2,-r_2) \Big<  b_{\omega }^{\dag}   b_{\omega' }\Big> \Bigg] \quad \text{for} \quad  d=2
 \end{gather}
 and
\begin{gather}
\nonumber
W_\beta\Big(t_1-t_2,\, r_1,r_2, \Omega_1,\Omega_2\Big) = \Big< {\phi}(t_1,r_1,\Omega_1){\phi}(t_2,r_2,\Omega_2) \Big> =\\= \sum_{k,k',j,j'}\int_{0}^{\infty}d\omega  d\omega'   \Bigg[\phi_{\omega j k}(t_1,r,\Omega_2)\phi^*_{\omega' j' k'}(t_2,r_2,\Omega_2) \Big<  a_{\omega j k}  a_{\omega' j' k'}^{\dag} \Big>
\nonumber
\label{wightman0}
+\\+
\phi^*_{\omega j k}(t_1,r,\Omega_2)\phi_{\omega' j' k'}(t_2,r_2,\Omega_2) \Big<  a_{\omega j k}^{\dag}   a_{\omega' j' k'}\Big> \Bigg] \quad \text{for} \quad  d>2,
 \end{gather}
where
$$
\Big<  a_{\omega }^{\dag}   a_{\omega' }\Big>    = n(\omega)\delta(\omega-\omega'),\quad \Big<  b_{\omega }^{\dag}   b_{\omega' }\Big>   = n(\omega)\delta(\omega-\omega'),
$$
$$\Big<  a_{\omega j k}^{\dag}   a_{\omega' j' k'}\Big>   = n(\omega)\delta(\omega-\omega')\delta_{j j'}\delta_{k k'}, \quad {\rm and} \quad n(\omega)= \frac{1}{e^{\beta \omega }-1}
$$
--- is the  Bose-Einstein or Planckian distribution. After some rearrangements, we obtain the following expressions for $d=2$:
\begin{gather}
\nonumber
  W_\beta(t_1-t_2,\, r_1,r_2)= \\= \int_{0}^{\infty}d\omega   \Bigg[
\Big(\phi^*_{\omega }(t_1,r_1)\phi_{\omega }(t_2,r_2) +\phi^*_{\omega }(t_1,-r_1)\phi_{\omega }(t_2,-r_2) \Big) \frac{1}{e^{\beta \omega }-1}
\nonumber
+\\+
\Big(\phi_{\omega}(t_1,r_1)\phi^*_{\omega }(t_2,r_2) +\phi_{\omega}(t_1,-r_1)\phi^*_{\omega }(t_2,-r_2) \Big) \left(1+\frac{1}{e^{\beta \omega }-1}\right)\Bigg].
\label{twopoinds2}
 \end{gather}
 and for $d>2$:
\begin{gather}
\nonumber
  W_\beta(t_1-t_2,\, r_1,r_2,\Omega_1,\Omega_2)= \\= \sum_{k,j}\int_{0}^{\infty}d\omega   \Bigg[
\phi^*_{\omega j k}(t_1,r_1,\Omega_1)\phi_{\omega j k}(t_2,r_2,\Omega_2) \frac{1}{e^{\beta \omega }-1}
\nonumber
+\\+
\phi_{\omega j k}(t_1,r_1,\Omega_1)\phi^*_{\omega j k}(t_2,r_2,\Omega_2)  \left(1+\frac{1}{e^{\beta \omega }-1}\right)\Bigg].
\label{twopoinds}
 \end{gather}
Using the thermal Wightman function, one can construct the thermal Feynman propagator as follows:
\begin{gather}
\nonumber
G_\beta (t_1-t_2,r_1,r_2,\Omega_1,\Omega_2)
=\\=
\theta(t_1-t_2) W_\beta (t_1-t_2,r_1,r_2,\Omega_1,\Omega_2)+\theta(t_2-t_1) W_\beta (t_2-t_1,r_2,r_1,\Omega_2,\Omega_1),
\end{gather}
which will be used below for the calculation of the effective action.

\section{Effective action in the 2d static de Sitter space-time}
\label{sec3}
In this section, we compute the one loop effective action for the massive scalar field in a two dimensional de Sitter space-time. And after the Wick rotation, we find the free energy. The thermal Feynman propagator in $2d$, which was found in the previous section, is as follows:
\begin{gather}
  \label{Green2d}
    G_\beta (t_1-t_2,r_1,r_2)=\int_{0}^{\infty} d\omega \frac{\sinh(\pi\omega)}{4 \cosh\pi(\nu-\omega)\cosh\pi(\nu+\omega)}
    \times  \\ \times
\nonumber
\Bigg[
   \frac{1}{e^{\beta \omega}-1}   e^{i\omega |t_1-t_2|} \left( \mathsf{P}^{i \omega}_{-\frac12 +i \nu}(r_1) \mathsf{P}^{-i \omega}_{-\frac12 +i \nu}(r_2)    + \mathsf{P}^{i \omega}_{-\frac12 +i \nu}(-r_1)
\mathsf{P}^{-i \omega}_{-\frac12 +i \nu}(-r_2) \right)
+\\+
\nonumber
 \left(1+ \frac{1}{e^{\beta \omega}-1}\right)   e^{-i\omega |t_1-t_2|} \left( \mathsf{P}^{-i \omega}_{-\frac12 +i \nu}(r_1) \mathsf{P}^{i \omega}_{-\frac12 +i \nu}(r_2)    + \mathsf{P}^{-i \omega}_{-\frac12 +i \nu}(-r_1)
\mathsf{P}^{i \omega}_{-\frac12 +i \nu}(-r_2) \right)\Bigg].
   \end{gather}
After the substitution of \eqref{Green2d} into \eqref{Seff}, one obtains that the effective action is:
  \begin{gather}
  \label{SPP}
\mathcal{S^\beta}_{\textsl{eff}}=
-\frac{T }{2}\int_{M^2}^{m^2} d\bar m^2 \int_{0}^\infty  d\omega
\frac{\sinh(\pi \omega)}{ \cosh\pi(\nu-\omega)\cosh\pi(\nu+\omega)} \left[\frac{1}{2}+\frac{1}{e^{\beta \omega}-1}\right]
\times\\ \times
\nonumber
 \int_{-1}^1 dr  \mathsf{P}^{i \omega}_{-\frac12 +i \nu}(r) \mathsf{P}^{-i \omega}_{-\frac12 +i \nu}(r),
\end{gather}
where $T$ is the duration of the Lorentzian time, and $M$ is very large in the curvature units.. The integral over $\omega$ in the last expression is divergent. This is the standard UV divergence due to the zero-point fluctuations. The situation here is absolutely similar to the one we have encountered in \eqref{FMIN} and \eqref{entRind}. Thus, we can simply renormalize this contribution as in the previous section.

To take the integral in (\ref{SPP}) over the spatial direction, $r$, we use the properties of the Legendre functions and the corresponding differential equation:
$$
\left[-\frac{\omega^2}{1-r^2} - \partial_r (1-r^2)\partial_r + m^2\right] \mathsf{P}^{i \omega}_{-\frac12 +i \nu}(r) = 0.
$$
Applying this equation, one can show what the integrand in (\ref{SPP}) can be written as a total derivative, and the answer for the $r$ integral in (\ref{SPP}) depends only on the behavior of the Legender functions at the limits of integration:
\begin{gather}
\nonumber
\int_{-1}^1 dr \, \mathsf{P}^{i \omega}_{-\frac12 +i \nu}(r) \mathsf{P}^{-i \omega}_{-\frac12 +i \nu}(r)
=\\=
\lim_{m'\rightarrow m} \int_{-1}^1  dr \, \frac{ \mathsf{P}^{-i \omega}_{-\frac12 +i \nu'}(r)\partial_r (1-r^2) \partial_r \mathsf{P}^{i \omega}_{-\frac12 +i \nu}(r)- \mathsf{P}^{i \omega}_{-\frac12 +i \nu}(r)\partial_r (1-r^2) \partial_r \mathsf{P}^{-i \omega}_{-\frac12 +i \nu'}(r)
}{m^2-m'^2}
\nonumber
= \\=
\label{PPINT}
\lim_{m\rightarrow m'} (1-r^2) \, \frac{\mathsf{P}^{-i \omega}_{-\frac12 +i \nu'}(r)\partial_r\mathsf{P}^{i \omega}_{-\frac12 +i \nu}(r)- \mathsf{P}^{i \omega}_{-\frac12 +i \nu}(r)\partial_r\mathsf{P}^{-i \omega}_{-\frac12 +i \nu'}(r)}{m^2-{m'}^2} \Bigg|_{-1}^{+  1}.
\end{gather}
Now, from the asymptotic behavior of the Legender functions, we find that:

\begin{gather}
\nonumber
\frac{ \sinh(\pi \omega)}{\cosh\pi(\nu-\omega)\cosh\pi(\nu+\omega)} \int_{-1}^1 dr \, \mathsf{P}^{i \omega}_{-\frac12 +i \nu}(r) \mathsf{P}^{-i \omega}_{-\frac12 -i \nu}(r)
= \\ =
\label{PPInt}
\frac{1}{2 \pi \nu }\left[ \psi\left(\frac{1}{2}-i \omega-i\nu \right)-\psi\left(\frac{1}{2}-i \omega+i\nu \right)+
\psi\left(\frac{1}{2}+i \omega+i\nu \right)-\psi\left(\frac{1}{2}+i \omega-i\nu \right)\right].
\end{gather}
Where $ \psi(z)=\Gamma'(z)/\Gamma(z)$ is the digamma function with simple poles of the residue $-1$ at $z=-n$. After the substitution of \eqref{PPInt} into \eqref{SPP}, one can take the integral over $\bar{m}^2$ to find that:
\begin{align}
\label{SLog}
\mathcal{S^\beta}_{\textsl{eff}}= -\frac{iT}{2 \pi  } \int_{0}^{\infty} d\omega \frac{1}{e^{\beta \omega}-1} \,
   \log\left[\frac{\Gamma\left(\frac{1}{2}-i \omega-i\nu\right)\Gamma\left(\frac{1}{2}-i \omega+i\nu\right)}{\Gamma\left(\frac{1}{2}+i \omega+i\nu\right)\Gamma\left(\frac{1}{2}+i \omega-i\nu\right)}\right] - (\nu\rightarrow M).
\end{align}
Here $(\nu\rightarrow M)$ means the same term as explicitly written here, but with $\nu$ form (\ref{nunu}) replaced by $M$, which is assumed to be very large in comparison with the de Sitter curvature.

Now let us represent the thermal distribution as the sum of the geometric series to find that:

\begin{align}
\mathcal{S}^\beta_{\textsl{eff}}=-
 i T \frac{1}{2 \pi  }   \int_{0}^{\infty} d\omega  \sum_{n=1}^{\infty} e^{-\beta \omega n}
  \log\left[\frac{\Gamma\left(\frac{1}{2}-i \omega-i\nu\right)\Gamma\left(\frac{1}{2}-i \omega+i\nu\right)}{\Gamma\left(\frac{1}{2}+i \omega+i\nu\right)\Gamma\left(\frac{1}{2}+i \omega-i\nu\right)}\right] - (\nu\rightarrow M).
\end{align}
Taking the integral over $\omega$ by parts and using the relation:
\begin{align}
     \frac{e^{-\beta n |\omega|}}{\beta n}=\frac{1}{\pi}\int_{-\infty}^{\infty}dy \frac{ e^{i y \omega}}{y^2+(\beta n)^2},
\end{align}
one obtains that:
\begin{align}
\mathcal{S}^\beta_{\textsl{eff}}=-
    \frac{T}{2 \pi ^2 }   \int_{-\infty}^{\infty} d\omega  \sum_{n=1}^{\infty} \int_{-\infty}^{\infty}dy \frac{ e^{i y \omega}}{y^2+(\beta n)^2}
  \left[ \psi\left(\frac{1}{2}-i \omega-i\nu\right)+\psi\left(\frac{1}{2}+i \omega+i\nu\right)\right]  -(\nu\rightarrow M).
\end{align}
Here, one can take the integral over $\omega$ by closing the contour in the upper half of the complex $\omega$-plane, if $y>0$, and in the lower one, if $y<0$. The result is:
\begin{align}
\int_{-\infty}^{\infty} d\omega e^{i \omega y} \left[\psi\left(\frac{1}{2}-i \omega-i\nu\right)+\psi\left(\frac{1}{2}+i \omega+i\nu\right)\right]=-2 \pi  \sum_{k=0}^{\infty}e^{-i y \nu -\left(k+1/2\right)|y|}= -\pi  \frac{e^{-i y \nu }}{\sinh\frac{|y|}{2}}
\end{align}
Hence,
\begin{align}
\label{Sb2}
\mathcal{S}^\beta_{\textsl{eff}} =
   \frac{T}{2 \pi   }     \sum_{n=1}^{\infty}  \int_{-\infty}^{\infty}dy \frac{ 1}{y^2+(\beta n)^2}   \frac{e^{i y \nu }-e^{i y M }}{\sinh\frac{|y|}{2}}.
\end{align}
Thus, we arrive at the integral representation of the effective action. For the arbitrary value of $\beta$ and $\nu$, we don't know how to express the integral in \eqref{Sb2} via known special functions. As we will see in the next section, the explicit form for $\mathcal{S}^{\beta}_{\textsl{eff}}$ and its derivatives over $\beta$ can be found only for $\beta=2 \pi \, 2 ^n $, $n \in Z$.

By performing the Wick rotation and changing $T \to - i \beta$, due to the analytic properties of the Feynman propagator in the complex $t$-plane\footnote{They are essentially the same as in flat space-time.}, one finds the free energy:
\begin{align}
\label{Fb2}
iS^\beta_{eff}\rightarrow   -\beta F_\beta =
       \frac{\beta}{4 \pi   }  \int_{-\infty}^{\infty}dy \frac{\frac{\pi  y}{\beta} \coth \left(\frac{\pi  y}{\beta }\right)-1}{
     y^2}   \frac{e^{i y \nu }-e^{i y M }}{\sinh\frac{|y|}{2}},
   \end{align}
which can be used to find thermodynamic properties of the massive scalar field theory (or free Bose gas) in the de Sitter space-time. Similar integral representation was found earlier in recent works \cite{Anninos:2020hfj,Grewal:2021bsu}, were the author expressed the free energy in terms of the integral over the $SO(1,2)$ Harish-Chandra character.

The free energy (\ref{Fb2}) diverges in the limit $M\rightarrow \infty$. This divergence is the same as the one that appears in the Rindler coordinates and was discussed in the previous section. In fact, let us cut from the integration contour over $y$ a small $\epsilon$ vicinity of the origin, similarly to what we did in the previous section. After such a cut, one can take the limit $M\rightarrow \infty$ since the integrand in (\ref{Fb2}) is a fast oscillating function everywhere on such a contour. However, the resulting expression for the free energy diverges in the limit $\epsilon \to 0$ in the same way as it was the case in \eqref{EFFM}. In the remeining part of this section, we keep $M$ fixed since it is more convenient to regularize this way the free energy in the two-dimensional case.  Thus, for a finite value of $M$, the integrand in \eqref{Fb2} remains finite at the origin, $y=0$.

\subsection{Explicit free energy and entropy for $\beta=2\pi$ }
In this subsection, we show that the free energy \eqref{Fb2} and the corresponding entropy can be found explicitly when $\beta=2 \pi$. Since the entropy is defined as $\mathbf{S}=(\beta\partial_\beta -1)\log[Z] $, it is enough to find the value and the first derivative of the free energy for $\beta=2\pi$.

The free energy \eqref{Fb2} for $\beta=2\pi$ can be rewritten in the form:
\begin{align}
F_{2 \pi} = -\frac{1}{2\pi}\int_{0}^{\infty}dy \frac{\frac{y}{2} \coth \left(\frac{y}{2}\right)-1}{ y^2}   \frac{\cos( y \nu )-\cos(y M)}{\sinh\frac{y}{2}},
\end{align}
where we divided the region of integration in \eqref{Fb2} as $y \in (-\infty, 0]$ and $y \in [0,+\infty)$ and then changed $y \to - y$ in the first region.

To calculate this integral, let us represent it as the sum of two terms :
\begin{align}
\label{L1+L2}
F_{2 \pi}= \lim_{\epsilon\rightarrow 0}\left[L_1(\epsilon,\nu)+L_2(\epsilon,\nu)\right],
\end{align}
where:
\begin{align}
\label{L_1}
L_1 (\epsilon,\nu)=-\frac{1}{2\pi}\int_{\epsilon}^{\infty}dy \frac{\frac{1}{2} \coth \left(\frac{y}{2}\right)}{ y}   \frac{\cos( y \nu )-\cos(y M)}{\sinh\frac{y}{2}}=  -\frac{1}{4\pi}\int_\nu^M dm \int_{\epsilon}^{\infty}dy  \coth \left(\frac{y}{2}\right)  \frac{\sin( y m )}{\sinh\frac{y}{2}}
\end{align}
and
\begin{align}
\label{L_2}
L_2 (\epsilon,\nu)=\frac{1}{2\pi}\int_{\epsilon}^{\infty}dy \frac{1}{ y^2}   \frac{\cos( y \nu )-\cos(y M)}{\sinh\frac{y}{2}}=\frac{1}{2\pi}\int_\nu^M dm_1 \int^{m_1}_0 dm \int_{\epsilon}^{\infty}dy   \frac{\cos( y m)}{\sinh\frac{y}{2}}.
\end{align}
These two integrals diverge separately in the limit $\epsilon\rightarrow 0$, but the divergent contributions cancel each other in the sum \eqref{L1+L2}.Further on, we will use the table integral from \cite{Grad}:
\begin{gather}
   \label{F1int}
\int_\epsilon^\infty \frac{\exp (a x)}{\sinh ^b(x)} \, dx=\frac{2 e^{(a+1) \epsilon}}{(a+b) \,  \sinh^{b-1}(\epsilon)} \, _2F_1\left(1,\frac{2+a-b}{2} ;\frac{2+a+b}{2};e^{2 \epsilon}\right),
\end{gather}
where $_2F_1 $ is the hypergeometric function.
Using this relation, one can calculate the integrals over $y$ in \eqref{L_1} and \eqref{L_2}. Then, taking the limit $\epsilon\rightarrow 0$, we find:
   \begin{align}
   \label{F2pi}
F_{2 \pi} = \frac{1}{2 \pi}\Bigg[  \nu ^2 - 2 \nu \  \Re \left[i\psi ^{(-1)}\left(\frac{1}{2}-i \nu
   \right)\right]
   -4  \ \Re\left[ \psi ^{(-2)}\left(\frac{1}{2}-i \nu
   \right)\right]
   \Bigg]-(\nu\rightarrow M).
   \end{align}
Note that $F_{2 \pi}<0$ for all values of $m$, i.e., of $\nu$.

For large values of $m$, the leading contribution to the free energy is as follows:
\begin{eqnarray}
F_{2 \pi} \approx\frac{1}{12 \pi } \log \frac{m}{M}, \label{mM}
\end{eqnarray}
which is twice larger than the same contribution in the Rindler space-time \eqref{mMRind} for the canonical temperature $\beta = 2\pi$. To understand this point, let us consider the eigen-functions from the continuous spectrum of the well-known quantum mechanical scattering problem in the two-dimensional de Sitter space-time:
\begin{align}
\label{scat}
\bigg[- \partial_x^2 + \frac{m^2}{\cosh^2 x}\bigg] R_{\omega}(x)=\omega^2 R_{\omega}(x).
\end{align}
This equation follows from \eqref{Rdifeq} after the change of variables $\tanh(x)=r$.

In the limit of large mass the potential barrier in (\ref{scat}) gets large. Then, the main contribution to the effective action comes from the regions near both sides of the horizon (in $2d$), where the potential looks approximately the same as in the Rindler space, namely: $\frac{m^2}{\cosh^2 x}\rightarrow 4m^2 e^{- 2 x} $. Thus, each side of the horizon (at $r=\pm 1$) gives the same Rindler type contribution to \eqref{mM}. At the same time, in the limit of small mass, the free energy behaves as:
$$
F_{2 \pi} \approx\frac{1}{12 \pi } \log\frac{m^6}{M},
$$
where the expression under the logarithm is dimensionless since all the masses are measured in units of the de Sitter curvature, which we set to one. 

To find the entropy or heat capacity, one has to calculate the derivative of the free energy over $\beta$. Using the approach described above, one can calculate any derivative of the free energy with respect to $\beta$ at $\beta=2\pi$. However, we will calculate only the first derivative, which is relevant to our main discussion.  The derivative of \eqref{Fb2} with respect to $\beta$ is equal to:
\begin{align}
\partial_\beta F_{\beta=2\pi} =
       \frac{1}{2 \pi   }  \int_{0}^{\infty}dy\left[-\frac{\text{csch}^2\left(\frac{y}{2}\right)}{8 \pi }+\frac{\coth \left(\frac{y}{2}\right)}{4 \pi  y} \right]\frac{\cos( y \nu )-\cos(y M)}{\sinh\frac{y}{2}}.
\end{align}
Representing again this expression as the sum of two terms, as it was done in \eqref{L1+L2}, and using the integral \eqref{F1int}, one obtains that:
\begin{gather}
\nonumber
\partial_\beta F_{\beta=2\pi}
=
\frac{1}{32 \pi ^2} \Bigg[4 \nu ^2-2\left(1+4 \nu ^2\right)\Re\left[ \psi ^{(0)}\left(\frac{1}{2}-i \nu
   \right)\right]
   +\\+
   \label{dF2pi}
   16  \nu  \Re\left[i\psi ^{(-1)}\left(\frac{1}{2}-i \nu \right)\right]+16 \Re\left[ \psi ^{(-2)}\left(\frac{1}{2}-i \nu \right)\right]\Bigg] -(\nu\rightarrow M).
\end{gather}
As a result, the entropy $\mathbf{S} = \beta^2\partial_\beta  F_{\beta=2\pi}$ can be calculated exactly for any mass $\nu$. Obviously, the resulting expression contains the two-dimensional de Sitter analog of the Bekenstein-Hawking term. It can be singled out by taking the large mass limit. This contribution has such a form as $\mathbf{S}\approx -\frac{1}{3}\log\frac{m}{M}$. It is again twice larger than the Bekenstein-Hawking entropy in the Rindler space-time, due to the presence of two sides of the horizon in the two-dimensional de Sitter space-time.

Another interesting limit is that of small mass:  $\mathbf{S}\approx -\frac{1}{3}\log\frac{m^3}{M}$. Below we will see how the temperature $\beta$ is present in these limiting expressions.

\subsection{Explicit free energy and entropy for $\beta=2\pi 2^n$}
In this section, we explicitly find the free energy and entropy for $\beta=2 \pi 2^n$, $n=1,2,3,\dots$ in $2d$. Consider the difference of two free energies \eqref{Fb2} for temperatures differing by the factor of two:
\begin{align}
\label{rec1}
F_{ 2 \, \pi \, 2^{n+1}} -  F_{2 \, \pi \, 2^{n}}=\frac{1}{ 2^{n+3}\pi}\int_\nu^M  dm\int_0^\infty \frac{\tanh \left(2^{-n-2} x\right)}{  \sinh(\frac{x}{2})}  \sin (m x).
\end{align}
The integral in this recurrent relation can be written in a simpler form than in eq. \eqref{Fb2}. In fact, the integrand in \eqref{rec1} can be expressed in the form:
\begin{align}
\frac{\tanh\left(2^{-n-2} x\right)}{ \sinh \left(\frac{x}{2}\right) }=
   (-1)^{2^n+1}
   \frac{2^{-n-1}}{\cosh ^2\left(2^{-n-2} x\right)}
   +
   \sum _{s=0}^{n-1} \frac{1}{2^s}\sum _{k=1}^{2^{n-s}-1}
   \frac{\exp \left(\frac{x k}{2^{n+1}}\right) (-1)^{(2^s+k)}}{1+\exp
   \left(\frac{x}{2^{s+1}}\right)},
\end{align}
where we use the partial fractional expansion of the rational fraction. Then using the table integral as follows:
\begin{align}
\label{TabInt2}
\int_0^{\infty } \frac{\exp (a x)}{1+\exp (b x)} \, dx=\frac{\psi \left(1-\frac{a}{2 b}\right)-\psi
   \left(1-\frac{a+b}{2 b}\right)}{2 b}; \text{ if } \Re(a)<\Re(b) \text{ and } \Re(b)>0,
\end{align}
we get the recurrent relation of the form:
\begin{gather}
\nonumber
F_{2\pi   2^{n+1}} -  F_{2 \pi  2^{n}}=\Re\Bigg[\frac{(-1)^{2^n+1} 2^{-2-2 n}}{\pi }\Bigg( i 2^n \nu  \left[\psi ^{(-1)}\left(1/2-i 2^n \nu \right)-\psi
   ^{(-1)}\left(1-i 2^n \nu \right)\right]
   +\\+
      \label{RecRel}
   \psi ^{(-2)}\left(1/2-i 2^n \nu \right)-\psi ^{(-2)}\left(1-i
   2^n \nu \right)\Bigg)
   +\\+
   \nonumber
   \sum _{s=0}^{n-1} \sum
   _{k=1}^{2^{n-s}-1} \frac{(-1)^{2^s+k}}{2^{1+s}}\left[\psi ^{(-1)}\left(\frac{1}{2}-\frac{ 2^s k}{2^{1+n}}-i 2^s \nu \right)-\psi ^{(-1)}\left(1-\frac{2^s k}{2^{1+n}}-i 2^s \nu \right)\right]\Bigg] - (\nu \rightarrow M).
\end{gather}
Since the free energy for $\beta=2 \pi $ and the difference of the two expressions for free energy are known exactly from \eqref{F2pi}and \eqref{RecRel}, one can represent the free energy for $\beta=2 \pi  2 ^n $ as:
\begin{align}
F_{ 2\pi   2^{n}}=F_{ 2\pi }+\sum _{l=0}^{n-1}\left(F_{ 2\pi   2^{l+1}}-F_{ 2\pi   2^{l}}\right).
\end{align}
Furthermore, the recurrent relation for the first derivative of the free energy over $\beta$ is as follows:

\begin{gather}
\label{recb1}
\partial_\beta \left(\beta F_{\beta= 2\pi  \ 2^{n+1}}\right) -   \partial_\beta\left(\beta  F_{\beta=2 \pi \ 2^{n}}\right)
=\int_0^{\infty }\frac{\cos (\nu  y)-\cos (M y)}{2^{5+2 n} \pi  \sinh
   \left(\frac{y}{2}\right) \cosh ^2\left(2^{-2-n} y\right)} dy .
\end{gather}
Using the partial fractional expansion of the hyperbolic function in \eqref{recb1}, one can compute this integral as it was done above, and then solve the simple recurrent relation for arbitrary $n$. Namely, we use the following expansion:
\begin{gather}
\frac{1}{\sinh   \left(\frac{y}{2}\right) \cosh ^2\left(2^{-2-n} y\right)}
=\\=
\frac{1}{2^{n+1} \left(-1+e^{\frac{y}{2^{n+2}}}\right)}-\frac{1}{2^{n+1} \left(1+e^{\frac{y}{2^{n+2}}}\right)}+\frac{(-1)^{2^n} 2^{2-n}}{\left(1+e^{\frac{y}{2^{n+1}}}\right)^3}-\frac{3 (-1)^{2^n}
   2^{1-n}}{\left(1+e^{\frac{y}{2^{n+1}}}\right)^2}
   \nonumber
   -\\-
   \nonumber
   \frac{(-1)^{2^n} \left( -5+2^{1+2 n}\right)}{3\  2^{n} \left(1+e^{\frac{y}{2^{n+1}}}\right)}+
   \sum _{k=1}^n \sum _{s=1}^{2^k}\frac{(-1)^{2^{n-k}+s} 2^{1-n+k}
   \left(2^{k-1}-s+1\right)  \exp \left(\frac{y (s-1)}{2^{n+1}}\right)}{1+e^{\frac{y}{2^{n-k+1}}}}
\end{gather}
and obtain that:
\begin{gather}
\label{RecRel1}
\partial_\beta \left(\beta F_{\beta= 2\pi  \ 2^{n+1}}\right) -   \partial_\beta\left(\beta  F_{\beta=2 \pi \ 2^{n}}\right)
=\\=
\nonumber
-\frac{2^{-5-2 n}}{\pi } \Re\left[\psi \left(1-i 2^{2+n} \nu \right)\right]
-\frac{2^{-6-2 n} }{\pi }\Re\left[\psi \left(1-i 2^{1+n} \nu \right)-\psi \left(1/2-i 2^{1+n} \nu \right)\right]
+\\+
\nonumber
\frac{(-1)^{2^n} 2^{-6-2 n} \left(1+2^{1+2 n} \left(1+12 \nu ^2\right)\right)}{3 \pi } \Re\left[\psi\left(1/2-i 2^n \nu \right)-\psi \left(1-i 2^n \nu \right)\right]
   +\\+
   \nonumber
  \sum _{k=1}^n \sum _{s=1}^{2^k} \frac{(-1)^{2^{n-k}+s} }{2^{5+2n}\pi}
   \left(2^{k-1}-s+1\right)
\Re\left[\psi \left(\frac{1+2^{1+k}-i 2^{1+n} \nu -s}{2^{1+k} }\right)-\psi \left(\frac{1+2^k-i 2^{1+n} \nu -s}{2^{1+k}}\right)\right],
\end{gather}
where the table integral \eqref{TabInt2} was used.

Since the first derivative of the free energy \eqref{dF2pi} for the inverse temperature $\beta=2 \pi $ is known, one can represent the entropy for the inverse temperature $\beta=2 \pi  2 ^n $ as:
\begin{align}
\label{EntL}
\mathbf{S}_{2\pi   2^{n}}=2\pi   2^{n} \left[\sum _{l=0}^{n-1}\Big[ \partial_\beta \left(\beta F_{\beta= 2\pi  \ 2^{l+1}}\right) -   \partial_\beta\left(\beta  F_{\beta=2 \pi \ 2^{l}}\right)\Big] + 2\pi\partial_\beta F_{\beta= 2\pi} +F_{ 2\pi }- F_{2\pi   2^{n}}\right]
\end{align}
The same technique allows one to calculate explicitly any derivative of the free energy for $\beta=2 \pi  2 ^n $, but the corresponding expressions are rather complicated. We will not discuss them here.

\subsection{Explicit free energy and entropy for $\beta=\frac{2\pi}{ 2^n}$}
This section explicitly finds the free energy and entropy for the inverse temperatures of the form $\beta=\frac{2 \pi}{ 2^n}$, $n=1,2,3,\dots$ in $2d$. Here, we will use a bit different technique from the one used in the previous subsection. 

Let us again consider the difference of two free energies \eqref{Fb2} with temperatures differing by the factor of two:
\begin{gather}
\label{rec3}
F_{\frac{2\pi}{   2^{n+1}}} -  F_{\frac{2\pi}{   2^{n}}}
=\frac{2^{n-2}}{\pi}\int_\nu^M  dm\int_0^\infty  dy \frac{ \sin (m y)   }{\cosh \left(2^{n-1} y\right)   }\frac{\sinh \left(2^{n-1} y\right)}{ \text{sinh}\left(\frac{y}{2}\right)}
=\\=
\nonumber
\frac{2^{n-2}}{\pi} \int_\nu^M  dm\int_0^\infty  dy \frac{\sin (m y)  }{  \ \cosh \left(2^{n-1} y\right) } \sum _{k=-2^{n-1}}^{2^{n-1}-1}  e^{\frac{1}{2} (1+2 k) y} .
\end{gather}
Taking the integrals and solving the recurrent relation, one obtains that:
\begin{gather}
F_{\frac{2\pi}{   2^{n+1}}}
= F_{2\pi}
+\\+
\nonumber
\sum_{s=0}^{n-1} \sum _{k=-2^{s-1}}^{2^{s-1}-1} \frac{2^{s-2}}{\pi} \Re\left[\psi ^{(-1)}\left(\frac{ 1+2^s+2 k-2 i \nu }{2^{2+s}}\right)-\psi ^{(-1)}\left(\frac{1+3\ 2^s+2 k-2 i \nu }{2^{2+s}}\right)\right]-(\nu\rightarrow M).
\end{gather}
Similarly, the recurrent relation for the first derivative of the free energy over $\beta$ is as follows:
\begin{gather}
\label{recb4}
\partial_\beta \left(\beta F_{\beta=\frac{ 2\pi}  {\ 2^{n+1}}}\right) -   \partial_\beta\left(\beta  F_{\beta=\frac{2 \pi}{ \ 2^{n}}}\right)
=-\int_0^{\infty }\frac{\cos (\nu  y)-\cos (M y)}{2^{3-2 n} \pi  \sinh
   \left(\frac{y}{2}\right) \cosh ^2\left(2^{n-1} y\right)} dy
   =\\=
   \nonumber
   \frac{1}{16 \pi }
   \Re\Bigg[-\frac{2^{3+2 n}}{1+4 \nu ^2}
   +2^{1+2 n} \psi\left(-\frac{1}{2}-i \nu \right)
   +\\+
   \nonumber
\sum_{k=1}^{2^n}  \left[ \left(1+2^n-2 k+2 i \nu \right) \psi\left(\frac{1}{2}+  \frac{-1+2 k-2 i \nu }{2^{n+2}}\right)
   -\left(1+2^n-2 k+2 i \nu \right)
   \psi\left(\frac{-1+2 k-2 i \nu }{2^{n+2} }\right)\right]\Bigg].
\end{gather}
Since we know exactly the value of the first  derivative of the free energy \eqref{dF2pi} for $\beta=2 \pi $, we can represent the entropy for $\beta=\frac{2 \pi }{ 2 ^n }$ as:
\begin{align}
\label{EntH}
\mathbf{S}_{\frac{2 \pi }{ 2 ^n }}= \frac{2 \pi }{ 2 ^n } \left[\sum _{l=0}^{n-1}\Big[ \partial_\beta \left(\beta F_{\beta=\frac{ 2\pi}  {\ 2^{l+1}}}\right) -   \partial_\beta\left(\beta  F_{\beta=\frac{2 \pi}{ \ 2^{l}}}\right)\Big]+2\pi\partial_\beta F_{\beta= 2\pi} +F_{\beta= 2\pi }- F_{\beta= \frac{2 \pi }{ 2 ^n }}\right].
\end{align}
Note that any derivative of the free energy can be calculated for $\beta=\frac{2 \pi }{ 2 ^n }$, but we do not write it here explicitly.

Let us discuss now different limits of the thermodynamic quantities in question for the temperatures of the form $\beta= 2 \pi \, 2 ^n$, $n \in Z$. Namely, for large $m$, the leading contribution to the free energy is as follows:
$$
F_{2 \pi/2^n} \approx\frac{2^{2 n}}{12 \pi } \log \frac{m}{M}\approx \frac{\pi}{3 \beta^2} \log \frac{m}{M},
$$
which is twice larger than the same quantity in the Rindler space-time \eqref{EFFM} due to the presence of bifurcate horizons.  For small $m$ the lading contribution to free energy is:
\begin{align}
\label{2dlog}
F_{2 \pi/2^n} \approx\frac{2^n \log (m )}{2 \pi }-\frac{2^{2 n} \log (M)}{12 \pi }\approx \frac{1}{ \beta} \log m-\frac{\pi}{3 \beta^2} \log M.
\end{align}

The free energy \eqref{Fb2} in the low temperature limit can be represent as follows:
\begin{gather}
F_\beta \approx-
   \frac{1}{2 \pi   }     \sum_1^{\infty} \frac{ 1}{(\beta n)^2}   \int_{-\infty}^{\infty}dy   \frac{e^{i y \nu }-e^{i y M }}{\sinh\frac{|y|}{2}}=-
   \frac{\pi }{3 \beta ^2}2\Re\left[\psi \left(\frac{1}{2}-i \nu \right)-\psi \left(\frac{1}{2}-i M \right)\right],
\end{gather}
which can be also deduced from the previous equation.
Let us also point out here that the high temperature limit can only be investigated via the consideration of the exact expression for the free energy. The obtained expressions together with those found in the previous subsection allow one to find analytic expressions for the thermodynamic quantities in question for arbitrarily high and low (but discrete) temperatures.

\section{The case of $d > 2$ dimensions}
So far, we have considered the two-dimensional static de Sitter space-time. In this section, we extend our considerations to higher dimensions. It is shown in \cite{Akhmedov:2019esv} that the properties of in-out effective actions of the massive scalar field in the Poincaré and the global patches of the de Sitter space-time essentially depend on the dimension of space-time. In even space-time dimensions, there is an imaginary contribution, which hints on the particle creation, while in odd dimensions, the imaginary contribution is vanishing, which, however, does not mean that there is no particle creation in odd dimensions \cite{Akhmedov:2019esv}. In this paper, we consider thermal states in the in-in formalism in the static de Sitter patch. We will see that ``thermal effects'' also significantly depend on whether the dimension of the de Sitter space-time is even or odd.

By definition \eqref{Seff}, the effective action in the $d$ dimensional case can be written as:
\begin{gather}
\label{effDds}
\mathcal{S}^\beta_{\textsl{eff}}=-\frac{1}{2} \int_{M^2}^{m^2} d\bar m^2  \sum_{k,j}^{}\int_{0}^{\infty}\frac{d\omega}{ \omega}   \left[\frac{1}{2}+\frac{1}{e^{\beta \omega }-1}\right] \times
\\
\nonumber
\times \int d t \int_0^1 dr \  r^{d-2} R_{\omega j} (r)R^*_{\omega j}(r)\int d\Omega Y_{j k}(\Omega)Y^*_{j k}(\Omega).
\end{gather}
The $1/2$ term in the square brackets in the last expression leads to the divergence due to zero-point fluctuations. We treat it the same way as above. Furthermore, note that here, unlike the two-dimensional case, the presence of the cutoff $M$ is not sufficient to make the last expression finite: In higher than two dimensions the last expression is divergent. We cutoff this divergence by imposing $j < \Lambda$ for some large $\Lambda$ in the sum over $j$.

The integral over the spherical angles can be evaluated using the orthogonality condition:
\begin{align}
\int d\Omega Y_{j k}(\Omega)Y^*_{j' k'}(\Omega)=\delta_{j j'} \delta_{k k'}.
\end{align}
To evaluate the integral over the product of radial functions in (\ref{effDds}), we use the method of the previous sections, where we showed that the volume integral over the two modes can be written as a total derivative. Then the answer for the integral depends only on the behavior of the modes at the boundary of the patch --- near the horizon.

Using \eqref{Rdifeq}, one can reduce the integral over the product of the radial functions \eqref{Rmode} to:

\begin{gather}
\nonumber
\,\int_0^1 dr \  r^{d-2} R^*_{\omega l}(r,m)R_{\omega l}(r,m)
=\\=
\nonumber
\lim_{m'\rightarrow m}\int_0^1  dr  \frac{R_{\omega l}(r,m')^*\partial_r\left(r^2(1-r^2)\partial_r\right)R_{\omega l}(r,m)- R_{\omega l}(r,m)\partial_r\left(r^2(1-r^2)\partial_r\right)R^*_{\omega l}(r,m')}{m^2-m'^2}
= \\=
\nonumber
 \lim_{m'\rightarrow m} \frac{R_{\omega l}(r,m')^* r^2(1-r^2)\partial_r R_{\omega l}(r,m)-R_{\omega l}(r,m) r^2(1-r^2)\partial_r R_{\omega l}(r,m')^*}{m^2-{m'}^2} \Bigg|_0^1
=\\=
\nonumber
\frac{\omega}{4 \pi  \nu} \Bigg[
\psi \left(\frac{2 L-2 i \omega-2 i \nu+d-1}{4} \right)-\psi\left(\frac{2 L-2 i \omega+2 i \nu+d-1}{4}\right)
-\\-
\psi \left(\frac{2 L+2 i \omega-2 i \nu+d-1}{4}\right)+\psi\left(\frac{2 L+2 i \omega+2 i \nu +d-1}{4}\right)\Bigg].
\end{gather}
In all, taking the integral in \eqref{effDds} over the spatial volume and calculating the sum over $k$, we obtain at the following expression for the effective action:

\begin{gather}
\mathcal{S}^\beta_{\textsl{eff}}=\frac{T}{16 \pi} \sum_{j=0}^{ \Lambda} \frac{(j+d-4)!}{(d-3)! j!}(2j+d-3)\int_{M^2}^{m^2} \frac{d\bar m^2}{  \nu}  \int_{-\infty}^{\infty}d\omega   \frac{1}{e^{\beta \omega }-1} \nonumber
\times \\ \times \nonumber \Bigg[
\psi \left(\frac{1}{4} \left(2 j-2 i \omega-2 i \nu+d-1\right)\right)-\psi\left(\frac{1}{4} \left(2 j-2 i \omega+2 i \nu+d-1\right)\right)
-\\-
\psi \left(\frac{1}{4} \left(2 j+2 i \omega-2 i \nu+d-1\right)\right)+\psi\left(\frac{1}{4} \left(2 j+2 i \omega+2 i \nu +d-1\right)\right)\Bigg]. \label{Seff1}
\end{gather}
As one can see, we obtain almost the same result as in the two-dimensional case, \eqref{SPP} and \eqref{PPInt}, but with the additional sum over $j$, which gives the additional ultraviolet divergence that is cutoff by $\Lambda\to \infty$.

Performing the same manipulations with \eqref{Seff1} as in section \ref{sec3}, one obtains:
\begin{gather}
\mathcal{S}^\beta_{\textsl{eff}}=\frac{T}{2\pi} \sum_{j=0}^{ \Lambda}\frac{(j+d-4)!}{(d-3)! j!}(2j+d-3)\int_{-\infty}^\infty dy\sum_{n=1}^{\infty} \frac{e^{i \nu y}-e^{i M y}}{y^2+\beta^2 n^2}\frac{ e^{-|y|\left(\frac{d-3}{2}+j\right)}}{\sinh(|y|)}.
\end{gather}
In this expression we can cutoff from the region of integration over $y$ the interval of the width $\epsilon$ around the origin. Namely we assume that the integration range in the last expression is: $ \gamma=(-\infty , -\epsilon )\cup (\epsilon , \infty)$. This allows one to take the limits $\Lambda\to \infty$ and $M\to \infty$ since, over such a region, we integrate a fast oscillating function and the expression remains finite for $\epsilon \neq 0$. Taking the sum over $j$, we obtain the effective action of the following form:
\begin{align}
 \mathcal{S}^\beta_{\textsl{eff}}=\frac{T}{2^{d-1}\pi} \int_\gamma dy\sum_{n=1}^{\infty} \frac{1}{y^2+\beta^2 n^2}
\frac{e^{i \nu y}}{ \sinh ^{d-1}\left(\frac{|y|}{2}\right)}.
\end{align}
Performing the summation over $n$ in the last expression, one finds:
\begin{align}
 \mathcal{S}^\beta_{\textsl{eff}}=\frac{T}{2^{d-1}\pi} \int_\gamma dy\frac{\frac{\pi  y }{\beta }\coth \left(\frac{\pi  y}{\beta }\right)-1}{2 y^2}
\frac{e^{i \nu y}}{ \sinh ^{d-1}\left(\frac{|y|}{2}\right)}.
\end{align}
After the analytical continuation to the imaginary time, one obtains the free energy of the form:
\begin{align}
\label{FD2}
 F_\beta=-\frac{1}{2^{d-1}\pi} \int_\gamma dy\frac{\frac{\pi  y }{\beta }\coth \left(\frac{\pi  y}{\beta }\right)-1}{2 y^2}\frac{e^{i \nu y}}{\sinh ^{d-1}\left(\frac{|y|}{2}\right)}.
\end{align}
The divergence of the integral at the origin, $y=0$, is cutoff by the choice of the contour $\gamma$. As we will see below the divergence of the resulting expression, as $\epsilon \to 0$, is the same as we have seen above in the Rindler chart. This integral representation was earlier found in a recent work \cite{Anninos:2020hfj}, where the authors expressed the free energy in terms of the integral over the $SO(1,d)$ Harish-Chandra character. We have learned about this result when the work on the present paper was finished.

From eq. \eqref{FD2}, one can see the difference between odd and even dimensions. That is due to the power of $\sinh ^{d-1}\left(\frac{|y|}{2}\right)$. In odd dimensions, the absolute value $|y|$ can be replaced by $y$ itself, and then one can use Cauchy's residue theorem to evaluate the integral. That explains the simplification in odd dimensions.

Furthermore, note that since the dimension of the de Sitter space-time manifests itself in the power of $\sinh ^{d-1}\left(\frac{|y|}{2}\right)$ in the denominator of the integrand in \eqref{FD2}, in any dimensions, one can do the same calculations as in the two-dimensional space-time. In particular, this means that in any dimension, one can calculate explicitly the value of the free energy and its derivatives for the discrete family of temperatures, $\beta= 2 \pi \, 2 ^n$, $n \in Z$. In this section, however, we restrict our considerations to the free energy properties only for odd dimensions and any $\beta$, since we can use Cauchy’s residue theorem to evaluate the integral  \eqref{FD2}.



Thus, adding a small semicircular contour $C_\epsilon$ around the origin $y=0$ and subtracting the same contribution and closing the contour by the infinite semicircular contour $C_R$, $R \to \infty$, we obtain that:
\begin{align}
F_\beta=\lim_{R\rightarrow \infty}\lim_{\epsilon\rightarrow 0}\left( I_{(-R , -\epsilon )\cup (\epsilon , R) } +I_{C_R}+ I_{C_\epsilon}-I_{C_\epsilon}\right).\label{IC}
\end{align}
Let us denote the sum of the first, second, and third terms as $F^{bulk}_\beta$ and the forth term as $F^{hor}_\beta$ since, as we will see below, these expressions can be interpreted as contributions from the bulk of the static de Sitter space-time and from the vicinity of the horizon correspondingly. The contour integrals in  $F^{bulk}_\beta$ give, via the Cauchy residue theorem, the expression of the form:
\begin{gather}
F^{bulk}_\beta=\lim_{R\rightarrow \infty}\lim_{\epsilon\rightarrow 0}
I_{(-R , -\epsilon )\cup C_\epsilon\cup (\epsilon , R) \cup C_R}
\nonumber =\\=2 \pi i \sum \text{Res} \left[ -\frac{1}{2^{d-1}\pi} \frac{\frac{\pi  y }{\beta }\coth \left(\frac{\pi  y}{\beta }\right)-1}{2 y^2}\frac{e^{i \nu y}}{\sinh ^{d-1}\left(\frac{y}{2}\right)}\right].
\end{gather}
Where the sum on the RHS of this expression runs over two sets of poles. The first set is due to the thermal distribution under the integral \eqref{FD2}. The positions of the poles in this set depend on the temperature, $\beta$. The second set is due to the poles of $\sinh ^{d-1}\left(\frac{y}{2}\right)$. Both sets together are:
\begin{align}
y=i \beta n , \quad n\in \mathbf{Z^+} \quad {\rm and} \quad  y=i 2\pi k \quad k\in \mathbf{Z^+}.
\end{align}
The fourth term in \eqref{IC} diverges in the limit $\epsilon\rightarrow 0$ and can be represented as:
\begin{align}
F^{hor}_\beta=-I_{C_\epsilon}=\sum_{k=1}^{\frac{d-1}{2}} \frac{a_{2 k-1}(m,\beta)}{\epsilon^{2k-1}} +a_0(m,\beta) +O(\epsilon), \label{coefexp}
\end{align}
where all $a_k$ are polynomial functions of the mass in the units of the de Sitter curvature. All terms in this expression, except $k=0$, can be absorbed into the renormalization of the gravitational effective action. The term with $a_0$ is equal to $ - \pi i$ times the residue at the origin. 

One can interpret $F^{hor}_\beta$ as the contribution from the boundary of the static patch --- from the horizon. First, recall that the divergent contributions are associated with the singularity of the thermal Feynman propagator at the horizon. Second, consider the limit of the large mass and small temperature. In such a case, the potential in the equation for the modes becomes exponentially large, as in the $2 d$ case \eqref{scat}. Then the main contribution to the effective action comes from the near-horizon region, as in the case of the Rindler coordinates. Hence, in such a limit, one of the coefficients in \eqref{coefexp}, which corresponds to the finite contribution in the limit $\epsilon\to 0$, acquires the form:
\begin{align}
\label{a0}
 a_0 (m,\beta)\approx \frac{(-1)^{\frac{d+1}{2}}}{3}\frac{m^{d-2}}{H \beta^2} \frac{A^{dS}_{d-2}}{2^d \pi^{\frac{d-2}{2}} \Gamma\left(\frac{d}{2}\right)},
\end{align}
where $ A^{dS}_{d-2}=\frac{2 \pi^{\frac{d-1}{2}}}{\Gamma\left(\frac{d-1}{2}\right)} \frac{1}{H^{d-2}}$ is the surface area of the boundary of the static de Sitter space-time. In this expression, we restored the Hubble constant $H$. This expression is consistent with the explicit form of the finite contribution to the effective action in the Rindler chart of flat space-time:
\begin{align}
F_{\beta}=\frac{(-1)^{\frac{d-1}{2}}}{3}\frac{m^{d-2}}{\alpha\beta^2} \frac{A_{d-2}}{2^d \pi^{\frac{d-2}{2}} \Gamma\left(\frac{d}{2}\right)}.
\end{align}
Where we also restored the proper acceleration $\alpha$. In the Rindler chart the finite part of the free energy contains only the contribution proportional to the area of the horizon divided by the acceleration: $F_{\beta}\sim\frac{A_{d-2}}{\alpha}$. As we see here, in the de Sitter space-time there is also similar contribution, which is proportional to the horizon area divided by the Hubble constant $a _0\sim\frac{A^{dS}_{d-2}}{H}$.


The contribution to the free energy, which cannot be attributed to the boundary, in odd dimensional case is as follows:
\begin{gather}
\label{FDb}
F^{bulk}_\beta= -\sum_{n=1}^\infty \frac{1}{2^{d-1} \beta  n }   \frac{ e^{- \beta  \nu n }}{\left[i \sin \left(\frac{n \beta }{2}\right)\right]^{d-1}}
-\\-
\nonumber
\frac{ i }{2^{d-2} (d-2)!}  \sum_{n=1}^\infty \frac{\partial^{d-2}}{\partial y^{d-2}} \frac{\frac{\pi  y }{\beta }\coth \left(\frac{\pi  y}{\beta }\right)-1}{2 y^2}\frac{\left(y- i 2 \pi n\right)^{d-1}}{\sinh ^{d-1}\left(\frac{y}{2}\right)} e ^{i \nu y} \Bigg|_{y= i 2 \pi n}.
\end{gather}
In the limit of large mass, the main contribution in the last expression comes from the the closest to the real axis pole, whose position depends on the temperature:
\begin{equation}
F^{bulk}_\beta\approx
 \begin{cases}
-\frac{ 1}{ 2^{d-1}\beta\left[i   \sin \left(\frac{ \beta }{2}\right)\right]^{d-1}}  e^{- \beta  m  } &,
 \text{if} \ \  \beta <  2 \pi \\ -\frac{(i m)^{d-2}}{  2^{d+1} \pi^2 i (d-2)!}
 \left[\frac{  2 \pi^2   }{\beta }\cot \left(\frac{2  \pi^2  }{\beta }\right)-1\right]  e^{- 2 \pi m}
 &, \text{if} \ \ \beta> 2 \pi
   \end{cases}.
\end{equation}
Thus, in the limit of large temperature, one obtains that $ F^{bulk}_\beta \sim e^{- \beta  m  } $, while in the low temperature limit --- $ F^{bulk}_\beta \sim e^{- 2 \pi  m  } $. I.e., the behavior of the leading contribution to the free energy changes at $\beta = 2 \pi$.

Let us consider also the situation in even dimensions.
Expanding the integrand in \eqref{FD2} around the origin one can represent the divergent (horizon) contribution to the free energy in even dimensions in a similar to (\ref{coefexp}) form: 
\begin{align}
F^{hor}_\beta\approx \sum_{k=1}^{\frac{d-1}{2}} \frac{b_{2 k}(m,\beta)}{\epsilon^{2k}} +b(m,\beta) \log(\epsilon)+O(\epsilon),
\end{align}
where $b_k$ are also polynomial functions of the mass measured in the units of the de Sitter curvature. Note that the divergent logarithmic term as $\epsilon \to 0$ is present only in even dimensions. At the same time, as we will see below, in the limit $\frac{m}{H}\to 0$ logarithmically divergent contributions appear in all dimensions.

The explicit form of $a_k$, $b_k$ coefficients is quite humongous and is not very informative. In fact, all $a_k$ and $b_k$ except $a_0, b_0$ do depend on the regularization scheme. Let us list a few relevant terms in small number of dimensions. E.g. the explicit form of the divergent contributions to the free energy in $d=3$ is as follows:  
\begin{align*}
F^{hor}_\beta\approx \frac{\pi }{3 \beta ^2 \epsilon },
\end{align*}
in $d=4$ it is:
\begin{align*}
F^{hor}_\beta\approx \frac{\pi }{6 \beta ^2 \epsilon ^2}+\frac{15 \pi  \beta ^2+60 \pi  \beta ^2 \nu ^2+8 \pi ^3 }{360 \beta ^4} \log (\epsilon ) ,
\end{align*}
and in $d=5$:
\begin{align*}
F^{hor}_\beta\approx \frac{\pi }{9 \beta ^2 \epsilon ^3}-\frac{\pi  \left(5 \beta ^2+15 \beta ^2 \nu ^2+2 \pi ^2\right)}{90 \beta ^4 \epsilon },
\end{align*}

Let us now restore the Hubble constant in \eqref{FDb} and consider the flat space limit $\frac{H}{m}\rightarrow 0$. In such a case $\nu=\sqrt{\frac{m^2}{H^2}-\left(\frac{d-1}{2}\right)^2}\approx \frac{m}{H}$. One can see that the second term in \eqref{FDb} vanishes in this limit because it contains exponentially suppressed terms: $e^{-2\pi n\frac{m}{H}}$. Hence, the leading contribution to the free energy is contained in the following expression:
\begin{align}
F^{bulk}_\beta\approx-\sum_{n=1}^\infty \frac{1} {2^{d-1}\beta  n } \frac{ e^{- \beta  m n }}{\left[i \sin \left(\frac{n \beta H }{2}\right)\right]^{d-1}} \approx -\sum_{n=1}^\infty \frac{1} {\beta  n } \frac{ e^{- \beta  m n }}{\left[i n \beta H \right]^{d-1}}.
\end{align}
Then, if $\beta m\rightarrow 0$, one obtains that:
\begin{align}
\label{limFds}
F^{bulk}_\beta\approx( -1)^{\frac{d+1}{2}}\frac{\zeta(d)} {H ^{d-1} \beta^d   }.
\end{align}
Using the spatial volume of the static de Sitter space-time: 	
\begin{align}
V^{sdS}_{d-1}=\int d \Omega_{d-2}\int_0^{1/H}d r \frac{r^{d-2}}{\sqrt{1-H^2 r^2}}=  \frac{\pi^{\frac{d}{2}}}{ H^{d-1}\Gamma\left(\frac{d}{2}\right)},
\end{align}
one can rewrite \eqref{limFds} as follows:
\begin{align}
F^{bulk}_\beta\approx  ( -1)^{\frac{d+1}{2}}  \frac{V^{sdS}_{d-1}}{\beta^d } \frac{\zeta(d) \Gamma\left(\frac{d}{2}\right)} {\pi^{\frac{d}{2}}    } .
\end{align}
This expression reproduces (up to the factor of $(-1)^{\frac{d+1}{2}}$) the one for the free energy of the massive scalar field in the Minkowski space-time \eqref{FMIN1}.

In all, the free energy in the static de Sitter space-time contains both $F^{hor}_\beta$, which contains divergent terms and is similar to the free energy in the Rindler chart, (these terms are proportional to the area of the horizon) and the finite $F^{bulk}_\beta$, which is proportional to the volume of the space-time and is similar to the free energy in the Minkowskian coordinates.

Now let us consider the free energy \eqref{FD2} in the limit of small mass (as compared to the Hubble constant) in the space-time of arbitrary dimension. In this limit:
\begin{align}
\nu=\sqrt{\frac{m^2}{H^2}-\frac{(d-1)^2}{2^2}}\approx -i\frac{d-1}{2}  +i \frac{ 1}{d-1}\frac{m^2}{H^2}.
\end{align}
As shown above, the integration over $y$ in \eqref{FD2} in the vicinity of the origin corresponds to the contribution from the boundary, and the rest of the integration region corresponds to the free energy of the bulk of the space-time.
To identify the main bulk contribution to the free energy in the limit of small mass, let us consider only the integration region from $y \in [L, \infty)$, where $L \gg 1/2$.  For such a contribution, one can use the following approximation for the denominator in \eqref{FD2}: $2^{d-1}\sinh ^{d-1}\left(\frac{|y|}{2}\right)\approx e^{\frac{d-1}{2}|y|} $. Hence, one obtains that:
\begin{align}
\label{logterm}
 F^{bulk}_\beta\approx-\frac{1}{2\beta}  \int_{L}^\infty \frac{dy}{y} e^{- \frac{ 1}{d-1}\frac{m^2}{H^2}  y}\approx \frac{1}{2 \beta} \log\left(\frac{L}{d-1}\frac{m^2}{H^2} \right)\approx\frac{1}{\beta}\log\left(\sqrt{L}\frac{m}{H} \right).
\end{align}
Thus, the main contribution to the bulk free energy in the limit $\frac{m}{H}\rightarrow 0$ contains a logarithmic term. Thus, we see that the logarithmic contributions to the free energy can appear also beyond $2d$ --- in any dimension.

\section{Conclusion and acknowledgments}
In this work, we consider the one-loop free energy and the entropy of the Gaussian massive scalar field theory in the static de Sitter space-time. We show that the divergent part of the free energy is associated with the additional (anomalous) singularities of the thermal Green functions on the horizon in the de Sitter space-time  \cite{Akhmedov:2020qxd, Akhmedov:2020ryq}.
We also consider the Rindler chart of the flat space-time as a model example.

In this work, we study in detail the two-dimensional de Sitter space-time. We find an analytic form for the free energy and the entropy for the family of inverse temperatures of the form $\beta=2 \pi 2^k; \ \ k\in \mathbf{Z}$. Also, in the odd dimensional de Sitter space-time for arbitrary temperatures, we separate contributions to the free energy that are divergent, and associated with the horizon, and that are finite, and are attributed to the interior of the space-time.

In the odd space-time dimension, we find the week field (large mass or small Hubble constant) expansion of the free energy. The leading contribution in this expansion depends on temperature: for large temperatures $ F_{\beta} \sim e^{- \beta  m  } $ and for low ones --- $ F_{\beta} \sim e^{- 2 \pi  m  } $. Thus, free Bose gas in the de Sitter space-time has certain properties that make it quite different from the one in the Minkowski space-time. 

We acknowledge useful discussions with K.V.Bazarov. This work was supported by the Foundation for the Advancement of Theoretical Physics and Mathematics “BASIS” grant, by RFBR grants 19-02-00815 and 21-52-52004, and by the Russian Ministry of Education and Science.

\bibliographystyle{unsrturl}
\bibliography{bibliography.bib}

\end{document}